\documentclass[english,onecolumn]{IEEEtran}
\usepackage[T1]{fontenc}
\usepackage{amsmath}
\usepackage{amsthm}
\usepackage{amssymb}
\usepackage{graphicx}
\usepackage{setspace}
\makeatletter
\theoremstyle{definition}
\ifx\thechapter\undefined
  \newtheorem{defn}{\protect\definitionname}
\else
  \newtheorem{defn}{\protect\definitionname}[chapter]
\fi
\theoremstyle{plain}
\ifx\thechapter\undefined
  \newtheorem{thm}{\protect\theoremname}
\else
  \newtheorem{thm}{\protect\theoremname}[chapter]
\fi
\theoremstyle{plain}
\ifx\thechapter\undefined
  \newtheorem{lem}{\protect\lemmaname}
\else
  \newtheorem{lem}{\protect\lemmaname}[chapter]
\fi
\theoremstyle{plain}
\ifx\thechapter\undefined
  \newtheorem{prop}{\protect\propositionname}
\else
  \newtheorem{prop}{\protect\propositionname}[chapter]
\fi

\usepackage[figurename=Fig.]{caption}
\usepackage[caption=false,font=footnotesize]{subfig}
\usepackage{cite}

\makeatother

\usepackage{babel}
\providecommand{\definitionname}{Definition}
\providecommand{\lemmaname}{Lemma}
\providecommand{\propositionname}{Proposition}
\providecommand{\theoremname}{Theorem}

\begin{document}

\title{Approximate Best-Response Dynamics in Random Interference Games}

\author{Ilai~Bistritz,~\IEEEmembership{Student Member,~IEEE}, and Amir~Leshem,~\IEEEmembership{Senior Member,~IEEE}\thanks{Ilai Bistritz is with the department of Electrical Engineering-Systems,
Tel-Aviv University, Israel, e-mail: ilaibist@gmail.com.

Amir Leshem~is with the Faculty of Engineering, Bar-Ilan University,
Ramat-Gan, Israel, e-mail: leshema@eng.biu.ac.il.

This research was supported by the Israel Science Foundation, under
grant 903/2013, and partially supported by the Israeli ministry of
science grant no. 204354. \protect \\
Parts of this paper were presented at the 55th IEEE Conference on
Decision and Control \cite{Bistritz2016}.}}
\maketitle
\begin{abstract}
In this paper, we develop a novel approach to the convergence of Best-Response
Dynamics for the family of interference games. Interference games
represent the fundamental resource allocation conflict between users
of the radio spectrum. In contrast to congestion games, interference
games are generally not potential games. Therefore, proving the convergence
of the best-response dynamics to a Nash equilibrium in these games
requires new techniques. We suggest a model for random interference
games, based on the long term fading governed by the players' geometry.
Our goal is to prove convergence of the approximate best-response
dynamics with high probability with respect to the randomized game.
We embrace the asynchronous model in which the acting player is chosen
at each stage at random. In our approximate best-response dynamics,
the action of a deviating player is chosen at random among all the
approximately best ones. We show that with high probability, with
respect to the players' geometry and asymptotically with the number
of players, each action increases the expected social-welfare (sum
of achievable rates). Hence, the induced sum-rate process is a submartingale.
Based on the Martingale Convergence Theorem, we prove convergence
of the strategy profile to an approximate Nash equilibrium with good
performance for asymptotically almost all interference games. We use
the Markovity of the induced sum-rate process to provide probabilistic
bounds on the convergence time. Finally, we demonstrate our results
in simulated examples.
\end{abstract}


\IEEEpeerreviewmaketitle{}

\section{Introduction}

\IEEEPARstart{A}{pplications} of game theory to networks in general
and wireless networks in particular is an active field of research
(see \cite{Han2012,Larsson2009,Saad2012,Leshem2009,Laufer2005}).
By modeling each user or node on a wireless network (e.g. mobile device,
access point, etc.) as a player, many conflicts and dynamics that
occur in wireless networks can be modeled as a game and analyzed using
game theoretic tools. Beyond pure analytic benefits, the game theoretic
tools can contribute to the development of distributed algorithms,
since each player acts independently and responds to his environment.
The outcome of the interaction between these independent nodes is
predicted by game-theoretic solution concepts. The best known solution
concept is the Nash equilibrium (NE, see \cite{Nash1951}). An analysis
of the NE points of a game can indicate how good the outcome of the
interaction is, in terms of designer-defined performance.

When attempting to apply game theoretic tools for engineering and
algorithmic purposes, the question of convergence to an equilibrium
is crucial, and goes beyond simply justifying that some equilibrium
notion can predict the outcome of a game well. No equilibrium can
have a practical meaning without being a natural stable point for
some dynamics that are sufficiently easy to implement.

Best-Response Dynamics (BR Dynamics) are well known in game theory.
In BR dynamics each player plays his best response to the current
actions of the other players. This means that the NE is a natural
stable point for BR dynamics. Due to their distributed nature and
simple implementation, they were widely applied for various tasks
over networks \cite{Altman2009a,Arcaute2009,Gharehshiran2014}. Simple
as they are, they are not guaranteed to converge.

In \cite{Goemans2005}, the authors presented the concept of sink
equilibria, which is a set of strategy profiles that pure BR dynamics
cannot escape from, once they attain it. While convergence of randomized
pure BR dynamics to a sink equilibrium is guaranteed, the result in
general is an oscillating dynamics which is impractical and undesirable
in many applications. Furthermore, without any additional information
it is impossible to analyze the performance of the resulting equilibria
or the convergence rate.

A necessary condition for BR dynamics to converge to a pure NE (PNE)
from any initial strategy profile is that there exists a path induced
by BR moves that connects the initial strategy to a PNE. A game with
such a property is called a weakly acyclic game \cite{Fabrikant2010a}.
In some scenarios, this condition can be considered sufficient, since
it has been shown that randomized BR dynamics converge almost surely
in weakly acyclic games \cite{Young1993} despite the fact that a
weakly acyclic game may contain cycles of BR moves.

While the convergence of BR dynamics in weakly acyclic games is indeed
encouraging, the main problem is that reliably checking weak acyclicity
is computationally intractable in the worst case \cite{Mirrokni2009}.
Other than that, it still does not provide indications as to the rate
of convergence to an equilibrium and cannot guarantee, in general,
the quality of the resulting NE.

A special case of weak acyclic games where more can be said about
convergence is potential games \cite{Monderer1996}. Potential games
are games in which there exists a function, known as the potential
function, such that for each player, any change in the utility causes
the same change (numerically or with the same sign) in the potential
function. Potential games exhibit at least one pure NE, and this NE
maximizes the potential function. Furthermore, a sequential BR dynamics
can easily be shown to converge to an NE. These are very desirable
properties, because if the potential function has something to do
with the performance required, they indicate that the convergence
is to a good NE \cite{Marden2009}. Motivated by these appealing properties,
many game formulations distributed algorithms in networks have used
potential games (see \cite{Altman2009,Cohen2013,Scutari2006,Cohen2017,Cortes2015,Marden2013}).

Congestion games are a general model for resource allocation games,
and are a special class of potential games. In a congestion game,
there are some available routes from a source to a destination, and
each player's strategy is to choose his own routes from the set of
all routes. The utility function for the player is the sum of the
delays in his chosen routes, where the delay in a route is a function
that decreases with the number of players who chose this route. It
has been shown that convergence to a pure NE in congestion games can
be very slow \cite{Fabrikant2004}. In fact, there are examples of
games and initial strategies where the shortest path to an equilibrium
in the BR dynamics is exponentially long in terms of the number of
players. Fortunately, if one is willing to settle for an approximate
NE, convergence can be a lot faster \cite{Chien2007}.

While congestion games have successfully been used to model resource
allocation in networks \cite{Cesana2008,Anshelevich2008,Marden2014},
they represent a high-layered approach and are not suited to modeling
similar scenarios in the physical layer. To the best of our knowledge,
the only effort that has been made to generalize congestion games
to wireless networks scenarios is \cite{Tekin2012}. The authors added
an interaction graph that defines which player is interacting with
whom, and so captures some notion of spatial geometry. Although constituting
a step forward, convergence to an NE was based again on a potential
function, which was shown only for special cases of interaction graphs.
Moreover, the interaction graph does not really capture the continuous
nature of the interference in a wireless network. From a more general
perspective, the authors in \cite{Candogan2013,Candogan2013a} studied
the properties of games that are near-potential in terms of their
utility functions, and suggested dynamics that converge to an approximate
NE for such games. The accuracy of the approximate NE is dependent
on the distance of the game from a potential game.

The family of interference games appears naturally when modeling the
capacity of wireless networks and frequency allocation problems. Here
we aim to generalize the desired convergence properties of congestion
games to these games. This is not an easy task. First, interference
games tend not to be potential games, and so BR dynamics convergence,
if occurs, should be established using other techniques. Even worse,
there are examples of non-weakly acyclic interference games in which
the BR dynamics fail to converge \cite{Leshem2009}. Our goal is to
overcome these obstacles by analyzing an approximate BR convergence
for a random interference game, where the channel gains represent
the long term fading determined by the players' locations. We show
that, asymptotically in the number of players, almost all interference
games do have converging approximate BR dynamics.

We analyze the probability for convergence asymptotically in the number
of players. The area of the region where players are located is kept
constant. The large number of players limit is toward denser networks
with a large number of devices in a given area, and has special significance
due the recent growing interest in large-scale networks. Nevertheless,
this limit has analytical advantages (for example, see \cite{Altman2002}).
Two closely related branches of game theory that exploit this limit
regularly are mean-field games and evolutionary games. In mean-field
games, each player is acting in response to an aggregated state created
by the choices of the mass \cite{Grammatico2014,Tembine2014}. Our
approach is essentially different from the above approach since in
our game the influence of each player on specific other players can
be fatal, and not negligible. In evolutionary games the analyzed dynamics
are of the relative parts of the population that chose each strategy
\cite{Sandholm2010,Altman2010,Madeo2015}. This requires that players
will be anonymous - only the number of players choosing a specific
strategy matters. In contrast, in interference games each player has
a unique characteristic that stems from his position in space. In
this sense, our results can be thought of as an alternative approach
to large-scale games that maintains the individuality and uniqueness
of each player. Another essential issue is that the above approaches
require a large number of players to be considered valid. In our approach,
the percentage of games for which the results are valid grows as the
number of players grows, and can be bounded from below for each finite
number of players.

Together with the assumption of fixed players' locations during the
dynamics, our approach is best suited for large scale communication
scenarios such as the internet of things (IoT), wireless sensor and
actuator networks (WSAN), cellular backbone and infrastructure with
picocells and femtocells, smart homes and cities, hotspots and networks
with pedestrian mobility.

The rest of this paper is organized as follows. In Section II we define
interference games and formulate the random interference game and
our system model. In section III we define our Approximate BR Dynamics,
and in Section IV we prove that these dynamics induce a sum-rate process
that is a submartingale. This establishes the convergence of the dynamics
to an approximate NE based on the Martingale Convergence Theorem.
Section V provides an analysis of the convergence time of the dynamics,
and Section VI presents simulations that confirm our results. Finally,
we draw conclusions in section VII.

This paper provides full proofs for the results presented in \cite{Bistritz2016},
along with a much more detailed discussion. Additionally, this paper
generalizes several assumptions made in \cite{Bistritz2016} (like
constant load, omnidirectional transmission and the scheduling) and
adds a convergence time analysis. The numerical results have also
been extended.

\section{Interference Games}

In this section we introduce interference games. We define a random
model for interference games and specify our system model in the limit
of a large number of players.

The fundamental conflict that arises from interference in the physical
layer has become a source for many game-theoretic formulations (see
\cite{Leshem2008,Larsson2009,Bistritz2015,Chung2003,Chen2012,Leshem2009,Laufer2005}
and the references therein), many of them use the transmission power
as the strategy of each player. Here we aim to define a general framework
for channel selection games in the interference channel, and argue
that in this form they are a natural generalization to the well-known
congestion games. All the games we deal with have a finite number
of players and finite strategy spaces.

We start by introducing congestion games, which are a special case
of potential games which is useful for modeling resource allocation
scenarios.

Throughout this paper, we use the standard game-theoretic notation
where $\boldsymbol{a}_{-n}=\left(a_{1},...,a_{n-1},a_{n+1},...,a_{N}\right)$
denotes the strategy profile of all player $n$'s opponents, and $A_{-n}=A_{1}\times...\times A_{n-1}\times A_{n+1}\times...\times A_{N}$
is the space of all those strategy profiles.
\begin{defn}[Ordinal Potential Game]
 A normal-form game $G=<\mathcal{N},\left\{ A_{n}\right\} _{n\in\mathcal{N}},\left\{ u_{n}\right\} _{n\in\mathcal{N}}>$
is an ordinal potential game if there exists a function $\Phi:A_{1}\times...\times A_{N}\rightarrow\mathbb{R}$
such that for each $n\in\mathcal{N}$ and for all $a_{n},a_{n}'\in A_{n}$
and $\mathbf{a}_{-n}\in A_{-n}$ , $\Phi(a_{n},\mathbf{a}_{-n})-\Phi(a_{n}',\mathbf{a}_{-n})>0$
if and only if $u_{n}(a_{n},\mathbf{a}_{-n})-u_{n}(a_{n}',\mathbf{a}_{-n})>0$.
\end{defn}
A potential game has at least one pure Nash equilibrium \cite{Rosenthal1973}.
If the game has not yet reached an equilibrium, then the turn of a
deviating player will arrive eventually and his action will increase
the potential function. After enough time, the potential function
will reach a maximum and no player will benefit from deviating, so
convergence to an NE occurs.
\begin{defn}[Congestion Game]
 A game with $N$ players and a set of congestible elements $E$
is called a congestion game if for each $n$, $a_{n}\subseteq E$
for each $a_{n}\in A_{n}$ and the utility of player $n$ is in the
form $u_{n}\left(\mathbf{a}\right)=\sum_{e\in a_{n}}c_{e}\left(x_{e}\right)$,
where each $c_{e}$ is a positive monotonically decreasing function
and $x_{e}=\left|\left\{ n|e\in a_{n}\right\} \right|$.
\end{defn}
Next, we define interference games.
\begin{defn}[Interference Game]
 An interference game is a game with a set $\mathcal{N}$ of $N$
players (each is a transceiver), where each player $n$ has another
player $d\left(n\right)$ as his destination, and a set $\mathcal{K}$
of $K$ channels. Each player's action is a subset of channels, so
for each $n$, $a_{n}\subseteq\mathcal{K}$ for each $a_{n}\in A_{n}$.
The utility of each player is in the form $u_{n}\left(\mathbf{a}\right)=\sum_{k\in a_{n}}R_{k}\left(I_{k}\left(\mathbf{a}\right)\right)$,
where each $R_{k}$ is a positive monotonically decreasing function
and $I_{k}\left(\mathbf{a}\right)=\underset{\left\{ m|k\in a_{m}\right\} }{\sum}g_{m,d\left(n\right)}P_{m}$.
The coefficient $g_{m,d\left(n\right)}$ is the channel gain between
player $m$ to the destination of player $n$ , and $P_{m}$ is the
transmission power of player $m$.
\end{defn}
We emphasize that the definition above is general and allows each
player to choose multiple channels. This shows the analogy between
interference games and congestion games. However, for the sake of
simplicity, we focus in this paper on the case where each player chooses
a single channel among the set $\mathcal{K}$.

Interference games are essentially different from congestion games.
They generalize them in the sense that in an interference game, the
effect of some player $m$ on player $n$ is different for each different
choice of $n,m$. Although higher numbers of players produce higher
interference on average, the exact interference is a weighted sum
of the channel gains between the involved players. This fact, which
stems directly from the geometry of the players in space, has major
implications for the convergence of BR dynamics to a NE.

Interference games tend not to be potential games (see the example
in \cite{Bistritz2016}), and in fact most of them are not\footnote{This can be verified by detecting BR cycles.}.
Consequently, BR convergence is not guaranteed. Furthermore, one can
construct an interference game that is not weakly acyclic, so there
is no hope for BR dynamics to converge in all interference games \cite{Leshem2009}.
The same is true for an approximate NE with a small enough epsilon.
Alternatively, for any epsilon there is an interference game with
no such approximate NE. Nevertheless, we are interested in typical
scenarios and the behavior of the majority of interference scenarios.
Theoretically, each interference game realization, i.e. a set of channel
gains, may have different equilibria and convergence properties, and
there is an infinite number of such realizations. We use the following
random model to show that, asymptotically in the number of players,
almost all of them do have converging (approximate) BR dynamics.

\subsection{Random Interference Game}

It is a common practice in wireless networks to model the channel
gains as random variables with some appropriate distribution. Our
goal is to analyze the probability that a certain interference game,
drawn at random according to this distribution, will have converging
approximate BR dynamics. In this subsection, we propose a random model
for interference games. It is by no means the only reasonable random
model, and other scenarios may require different models.

Our channel gains distribution is dictated by the random player locations.
We assume that the players' locations are generated uniformly and
independently at random in some subregion of the real plane, $\mathcal{D}\subseteq\mathbb{R}^{2}$,
with area $\left|\mathcal{D}\right|$. Each player as a transmitter
has some other player's receiver as his destination. We also explore
the effect of beamforming on the convergence.
\begin{defn}[Channel Gains]
 \label{Channel Gains}Assume that each transceiver is capable of
transmitting a beam with angle $\theta_{T}$ and receiving a beam
with angle $\theta_{R}$, with some arbitrary orientation. The channel
gains are determined directly from the players' locations and the
beamforming angles, so the channel gain between player $n$ to the
destination of player $m$ is defined by
\begin{equation}
g_{n,d\left(m\right)}=\frac{G}{r_{n,d\left(m\right)}^{\alpha}}I_{d}\left\{ \left|\theta_{n,d\left(m\right)}\right|\leq\frac{\theta_{T}}{2}\right\} I_{d}\left\{ \left|\theta_{d\left(m\right),n}\right|\leq\frac{\theta_{R}}{2}\right\} \label{eq:1-1}
\end{equation}
where $I_{d}$ is the indicator function, $r_{n,d\left(m\right)}$
is the Euclidean distance between player $n$ and player $m$'s destination,
$\alpha$ is the path-loss exponent and $G=\left(\frac{\nu}{4\pi}\right)^{\alpha}$,
where $\nu$ is the wavelength. The angles $\theta_{n,d\left(m\right)}$
and $\theta_{d\left(m\right),n}$ are defined with respect to the
beamforming angle bisectors of players $n$ and $d\left(m\right)$,
respectively.
\end{defn}
The beamforming angels are chosen such that the interference (transmitted
or received) outside the beam can be neglected (e.g., 10 dB beamwidth).
This type of beamforming is natural for large scale networks, since
the number of interferers tend to be larger than the degrees of freedom
of the multiple-input and multiple-output (MIMO) transceiver, so zero
forcing is impractical. Acquiring the required channel state information
in a distributed network is also a major issue. Substituting $\theta_{T}=\theta_{R}=2\pi$
in \eqref{eq:1-1} results in the special case of omnidirectional
transmission and reception.

Our model does not include instantaneous small scale-fading. This
may be considered a result of averaging the channel gains over multiple
coherence times, which occurs if our dynamics have a duration that
is significantly larger than the coherence-time of the wireless channel.
From a system perspective this is very natural, since each player
needs to sense the interference in each frequency band in order to
choose his new channel; hence channel switching typically requires
a longer time scale than the coherence time. Additionally, the assumption
of constant locations is better suited to a scenario with mobility
rates that are much slower than the convergence time of the dynamics.

We use the achievable rate when interference is treated as noise as
our utility function
\begin{equation}
u_{n}\left(\mathbf{a}\right)=R_{n}\left(\mathbf{a}\right)=R_{n}\left(I_{n}\left(\mathbf{a}\right)\right)=\log_{2}\left(1+\frac{g_{n,d\left(n\right)}P_{n}}{N_{0}+I_{n}\left(\mathbf{a}\right)}\right)\label{eq:2}
\end{equation}
where $P_{n}$ is player $n$'s transmission power, which is limited
by $P_{n}\leq P_{\textrm{max}}$ for all $n$, and $N_{0}$ is the
Gaussian noise variance. We emphasize that this utility is of player
$n$ and does not reflect on the utility of player $d\left(n\right)$.
The utility of player $m=d\left(n\right)$ is measured by his achievable
rate in the transmission to $d\left(m\right)$. Observe that the social
welfare of an interference game is the sum of the achievable rates,
which is a desirable performance measure for a channel allocation
algorithm. We assume that each player can sense the exact interference
in each of the $K$ channels and adapts his transmission scheme accordingly.
This sensing capability allows each player to compute his utility
function in each channel, provided that other players stay in their
current channel.

\subsection{System Model}

As in ad-hoc networks, in our network each player has another player
as his destination. This need not be his final destination, but just
the next hop. We assume that these destinations are constant over
the duration of our dynamics, and that each formed link is constantly
active (although that not necessarily with a continuous transmission).

Our analysis is asymptotic in $N$, the number of players. A similar
approach was taken in \cite{Bistritz2015}, but for analyzing the
structure of Nash equilibria in frequency-selective interference games.
The goal of this subsection is to define how the system parameters
behave in the limit where $N$ approaches infinity. These parameters
are the number of channels, the transmission powers and the area of
the players' region. Note that since each player demands some specific
target rate, the transmission power of each player is dictated by
his distance from his destination and vice versa. A parameter that
is fixed with respect to $N$ will be termed ``constant''.

We assume that the region $\mathcal{D}\subseteq\mathbb{R}^{2}$, where
the players are located, has area $\left|\mathcal{D}\right|=\frac{1}{\lambda}$
for some $\lambda>0$, so it is fixed with respect to $N$. This yields
an increasing players' density as $N\rightarrow\infty$. This is in
the spirit of large scale networks such as the IoT, sensor and actuator
networks, cellular networks with femtocells, smart homes and cities
and others.

Now we have to decide about $K$, the number of channels, and $P_{\textrm{max}}$,
the maximal transmission power. We argue that the only reasonable
choice is having both $K\rightarrow\infty$ and $P_{\textrm{max}}\rightarrow0$.
If only $K\rightarrow\infty$ but $P_{\textrm{max}}$ is fixed with
respect to $N$, then there are a constant (or even an increasing)
number, $\frac{N}{K}$, of players per channel on average. A constant
$P_{\textrm{max}}$ allows each player to choose a destination from
all the other players in the region $\mathcal{D}$. Hence a player
with a destination that is within a constant distance from him will
suffer from $\frac{N}{K}-1$ strong interferers. Each of these interferers
transmits with a constant power and is within a constant distance
away. Consequently, this player will be limited by a negative (in
decibels) signal to interference ratio. Now assume instead that $K$
is fixed with respect to $N$ but $P_{\textrm{max}}\rightarrow0$.
In this case the average number of players per channel, $\frac{N}{K}$,
grows to infinity. Since $P_{\textrm{max}}\rightarrow0$ forces each
player to choose only a nearby player as a destination, each pair
of source-destination will be geometrically close (with a decreasing
distance with $N$). Hence, the growing number of players on a specific
channel have to be geometrically separated enough from each other.
Otherwise their signal to interference ratio will be negative (in
decibels) due to the aggregated interference. Unfortunately, this
is not possible, since the region $\mathcal{D}$ has a fixed area
with respect to $N$.

Following the argument above we assume that the system load per frequency,
defined as the ratio $l_{N}=\frac{N}{K}$, satisfies $l_{N}\left(\frac{\log N}{N}\right)^{\frac{\alpha}{\alpha+2}}\rightarrow0$
as $N\rightarrow\infty$, so $l_{N}=o\left(\left(\frac{N}{\log N}\right)^{\frac{\alpha}{\alpha+2}}\right)$
(where $\alpha$ is the path-loss exponent). This of course holds
for the special case of constant load, i.e., $l_{N}=l$ (presented
in \cite{Bistritz2016}), but even for $l_{N}=\sqrt{N}$ for any $\alpha>2$.
These assumptions yield an increasing (or at least constant) network
wide spectral efficiency in bps per Hz per square meter. Without loss
of generality, we assume that $\frac{N}{K}=l_{N}\geq1$, since the
analysis for all the cases where $N<K$ readily follows from that
of the $N=K$ case.

Interestingly, the requirement that $P_{\textrm{max}}\rightarrow0$
as $N\rightarrow\infty$ can be also deduced from a broader perspective,
that takes into account the routing mechanism applied in the network.
In the fundamental work \cite{Gupta2000}, it was shown that if the
(final) destinations are chosen arbitrarily in $\mathcal{D}$ then,
even if the users' locations, transmission ranges for each hop and
the transmission schedules are all centrally optimized, each user
will have a vanishing throughput when the number of users grows. This
result is valid under the assumptions of stationary users and interference
that is treated as noise. Dropping the stationarity assumption, it
has indeed been shown that mobility can be exploited to avoid the
vanishing throughput for the price of significant delays, but this
is far from practice for most applications \cite{Grossglauser2001}.
Instead of treating the interference as noise, advanced decoding techniques
such as interference cancellation or interference alignment can theoretically
be applied. Unfortunately, these are essentially impractical in large-scale
networks due to the immense channel state information required from
each user.

This leads us to assume that the transmission range of each player
is limited by $r_{\textrm{max}}=\sqrt{\frac{\log N+a_{N}}{\pi N}}$
for some constant $r_{0}>0$ and $a_{N}\rightarrow\infty$ as $N\rightarrow\infty$.
This transmission range is far from arbitrary in networks with random
and uniform players' placement. Define two players as connected if
they are within a transmission range from each other. The radius $r_{\textrm{max}}$
is known to asymptotically guarantee the connectivity of the network
\cite{Gupta1999}, and specifically allows a player to choose any
one of his nearest neighbors as the next hop. In fact, the throughput
increases as the transmission range decreases, and since the network
should maintain connectivity, $r_{\textrm{max}}$ is the ideal transmission
range. Indeed, for this choice, we obtain that if the number of hops
until the final destination, for each player, is constant with respect
to $N$, then the throughput per player does not vanish as $N\rightarrow\infty$.
If transmit or receive beamforming is used, the transmission range
that guarantees asymptotic connectivity remains the same up to a constant,
although a special care should be given to edge effects, antenna pattern
and the beamforming technique \cite{Diaz2003}. According to this
approach, a player that wishes to transmit packets to a far away player
should do so by choosing a wired nearby player as his physical destination.
This is possible if wired players are deployed densely enough and
the routing algorithm is designed accordingly.

To implement the assumption on the transmission range, we set the
maximal transmission power to $P_{\textrm{max}}=P_{0}\left(\frac{\log N}{N}\right)^{\frac{\alpha}{2}}$
for some $P_{0}>0$. If $r_{n,d\left(n\right)}\leq\sqrt{\frac{\log N}{N}}$
then $g_{n,d\left(n\right)}\geq G\left(\frac{N}{\log N}\right)^{\frac{\alpha}{2}}$
and so $P_{\textrm{max}}=P_{0}\left(\frac{\log N}{N}\right)^{\frac{\alpha}{2}}$
is enough to guarantee any target rate in the absence of interference.
Specifically, to guarantee an achievable rate of at least $R$ bps
in the absence of interference, one should choose $P_{0}=\frac{N_{0}}{G}\left(2^{R}-1\right)$.
In practice, almost all of the players will find a destination among
their nearest neighbors at a distance that scales like $\frac{1}{\sqrt{N}}$
rather than $\sqrt{\frac{\log N}{N}}$. Hence, such a player can use
power control and decrease his energy consumption by using only $P_{n}=\frac{P_{0}}{N^{\frac{\alpha}{2}}}$
and still obtain the required achievable rate.

Finally, we allow each player to be a destination to a maximum of
$S>0$ players, which is fixed with respect to $N$. This assumption
does not actually stand on its own. Rather it is naturally obtained
from the requirement that the next hop of each player is one of his
nearest neighbors since their number, on average, is constant with
respect to $N$.

Our model is a building block for many communication scenarios. For
instance, for duplex communication one can use parallel interference
games, one for each direction, where in the second game the transmitter
and receiver rules of each pair are reversed.

\begin{figure}[tbh]
~~~~~~~~~~~~~~~~~~~~~~~~~~~~~~~~~~~~~~~~~~~~~~~~\includegraphics[width=7cm,height=4.8cm]{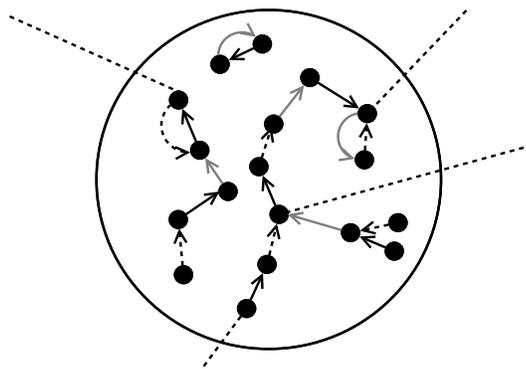}

\caption{\label{fig:Network}Illustration of the network for $N=18$ and $K=3$}
\end{figure}

A toy example of our network is depicted in Figure \ref{fig:Network}.
The destination of each player (dot) is shown with an arrow, and the
type of the arrow (dashed, solid or gray) represents one of three
available channels. Wired players are connected to the doted outgoing
lines.

An alternative limit to the one taken in this work is to let the area
of the region $\mathcal{D}$ grow linearly with $N$. This way the
number of channels $K$ could be kept constant as $N\rightarrow\infty$
without rendering the average throughput to vanish due to the interference.
In this limit the expected number of players in each channel goes
to infinity and the mean-field approach, or concentration results
such as \cite{Kiani2008}, can be applied to the interference.

\section{Approximate Asynchronous BR Dynamics}

In this section we define our approximate BR dynamics, which we show
are converging for asymptotically almost all interference games. The
idea behind our convergence result is that although our game is not
a congestion (or potential) game, it is a property of the typical
realization game rather than of the distribution of the random game.
While it is possible that some player may choose an action that decreases
the sum-rate, he has no statistical incentive to do so. A player simply
does not have enough information to choose an action that benefits
other players. This implies that if each player's action is random
enough, the average change in the sum-rate will be positive. Broadly
speaking, this means that the average game can be thought of as a
congestion game. Nevertheless, our dynamics should converge in almost
all realizations of networks and the fact that the average one behaves
well does not imply that most of them are.

Each realization possesses a large number of random variables: these
are the channel gains of the large number of players $N$. Therefore,
the statistics of the average action are in some sense similar to
those of any player in the average game, and all that is left is that
some mechanism will average over the actions and players. Fortunately,
this type of mechanism is very natural in reality. We suggest the
Approximate Asynchronous BR Dynamics to be this mechanism, as presented
in this section. The randomness of these dynamics has two sources:
the random acting player in each step, and the possibly random action
that he chooses.

Our time axis is discrete, with index $t$, and in each step some
random player chooses his action and the strategy profile changes.
We want $\left\{ \mathbf{a_{\mathbf{t}}}\right\} $ to converge to
some epsilon pure Nash equilibrium.
\begin{defn}
A strategy profile $(a_{n}^{*},\mathbf{a}_{-n}^{*})\in A_{1}\times...\times A_{N}$
is called an epsilon pure Nash equilibrium ($\varepsilon$-PNE) if
$u_{n}(a_{n}^{*},\mathbf{a}_{-n}^{*})\geq u_{n}(a_{n}\mathbf{a}_{-n}^{*})-\varepsilon$
for all $a_{n}\in A_{n}$ and all $n\in\mathcal{N}$.
\end{defn}
If every player acts asynchronously, the acting player in a given
step will emerge at random from some set of possible acting players.
This can be implemented by random channel access schemes, similar
in spirit to carrier-sense multiple access (CSMA) or the ALOHA protocol.
We emphasize that the transmission itself needs not be of short duration
as in the traditional schemes. In our dynamics, each player is using
his current chosen channel constantly, until an opportunity to switch
a channel arises.
\begin{defn}[Non-Degenerate Schedule]
Denote by $\mathcal{N}_{t}$ the set of players that can act on turn
$t$ and their number by $N_{t}$. In an asynchronous dynamics with
the schedule $\left\{ \mathcal{N}_{t}\right\} $, the acting player
on turn $t$ is chosen uniformly at random from $\mathcal{N}_{t}$.
We say that the schedule $\left\{ \mathcal{N}_{t}\right\} $ is non-degenerate,
with respect to the decision rules $h_{n}:\,A_{1}\times...\times A_{N}\rightarrow A_{1}\times...\times A_{N}$,
if for each $t\geq0$ either it contains at least one player $n$
such that $h_{n}\left(\mathbf{a}_{t}\right)\neq\mathbf{a}_{t}$ or
that $h_{n}\left(\mathbf{a}_{t}\right)=\mathbf{a}_{t}$ for each $n\in\mathcal{N}$.
\end{defn}
In a degenerate schedule, there could be turns where only players
who do not want to deviate have a chance to act. In a non-degenerate
schedule this can only happen in an equilibrium. A degenerate schedule
may lead to an infinite number of turns where nothing happens, which
makes the question of convergence not well defined. Natural choices
for a non-degenerate $\mathcal{N}_{t}$ are $\mathcal{N}_{t}=\mathcal{N}$
(all players) and $\mathcal{N}_{t}=\left\{ n\,|\,BR_{\varepsilon,n}\left(\mathbf{a}_{t}\right)\neq\mathbf{a}_{t}\right\} $
(the deviating players). For these choices of $\mathcal{N}_{t}$ each
player acts independently, and no centralized coordination is required.
Many other interesting choices can be based on fully distributed protocols.

Now we define our dynamics, which are to choose at random from the
set of $\frac{\varepsilon}{2}$-best actions, or keep the previous
action if it is currently $\varepsilon$-best. The randomness of the
action chosen by the acting player is a result of the indifference
between the approximately best actions. Our non-degeneracy of $\left\{ \mathcal{N}_{t}\right\} $
is with respect to these dynamics.
\begin{defn}
\label{eps-BR}Let $\varepsilon>0$. The set of $\varepsilon$-best
actions for player $n$ is defined as
\begin{equation}
B_{\varepsilon}\left(\mathbf{a}_{-n}\right)=\left\{ a\,|\,a\in A_{n},\,u_{n}(a,\mathbf{a}_{-n})+\varepsilon\geq\underset{a'\in A_{n}}{\max}u_{n}(a',\mathbf{a}_{-n})\right\} \label{eq:3-1}
\end{equation}
This is a finite and discrete set of channels. The $\varepsilon-$BR
response function of player $n$, $BR_{\varepsilon,n}\left(\mathbf{a,\omega}\right)\,:\,\omega\times A_{1}\times...\times A_{N}\rightarrow A_{1}\times...\times A_{N}$,
is defined as
\begin{equation}
BR_{\varepsilon,n}\left(\mathbf{a,\omega}\right)=\Biggl\{\begin{array}{cc}
\mathbf{a} & a_{n}\in B_{\varepsilon}\left(\mathbf{a}_{-n}\right)\\
(a_{\textrm{rand},n}\left(\omega\right),\mathbf{a}_{-n}) & o.w.
\end{array}\label{eq:3}
\end{equation}
where $a_{\textrm{rand},n}\left(\omega\right):B_{\frac{\varepsilon}{2}}\left(\mathbf{a}_{-n}\right)\rightarrow B_{\frac{\varepsilon}{2}}\left(\mathbf{a}_{-n}\right)$
is a random variable which is uniformly distributed over  $B_{\frac{\varepsilon}{2}}\left(\mathbf{a}_{-n}\right)$.
For simplicity, through the paper we use the notation $BR_{\varepsilon,n}\left(\mathbf{a}\right)$
instead, which is a random variable with distribution that is determined
by $\mathbf{a}$.
\end{defn}
In order to prove convergence, we have to find some function of $\mathbf{a_{\mathbf{t}}}$,
denoted by $X\left(\mathbf{a_{\mathbf{t}}}\right)$, for which the
expectation over the random actions grows with each step. This will
make $\left\{ X\left(\mathbf{a_{\mathbf{t}}}\right),\mathbf{a_{\mathbf{t}}}\right\} $
a submartingale (i.e., a submartingale with respect to the sigma algebra
generated by the sequence $\left\{ \mathbf{a_{\mathbf{t}}}\right\} )$.
The formalization of our proof is based on the martingale convergence
theorem \cite[Page 89, Theorem 5.14]{Breiman1992}. Let $X(\mathbf{a}_{t})$
be the sum of the achievable rates experienced by all players in their
chosen channel, i.e.,
\begin{equation}
X(\mathbf{a}_{t})=\sum_{n=1}^{N}R_{n}\left(\mathbf{a_{\mathbf{t}}}\right)\label{eq:4}
\end{equation}
This is a real non-negative bounded function of the strategy profile.
Recall that this is just the social welfare of an interference game.
This means that this convergent function is aligned with our centralized
objective and will guarantee that the resulting NE exhibits good performance.
To prove convergence we have to show that $E\left\{ X(\mathbf{\mathbf{a_{t+1}}})-X(\mathbf{a_{t}})|\mathbf{a_{t}}\right\} $
is a non-negative function of $\mathbf{a_{t}}$ for all $t\geq0$,
where the expectation is taken over the random acting player and the
random action that he may choose. Let $\mathbf{a_{t+1}}=BR_{\varepsilon,n}\left(\mathbf{a_{t}}\right)$,
so player $n$ is the acting player at time $t$ and observe that
\begin{equation}
X(\mathbf{a_{t+1}})-X(\mathbf{a_{t}})=\varDelta_{1}\left(\mathbf{a_{t}}\right)+\varDelta_{2}\left(\mathbf{a_{t}}\right)+\varDelta_{3}\left(\mathbf{a_{t}}\right)\label{eq:5}
\end{equation}
where
\begin{equation}
\varDelta_{1}\left(\mathbf{a_{t}}\right)=R_{n}\left(\mathbf{a_{\mathbf{t}+1}}\right)-R_{n}\left(\mathbf{a_{\mathbf{t}}}\right)\label{eq:5-1}
\end{equation}
is the rate difference of the acting player,
\begin{equation}
\varDelta_{2}\left(\mathbf{a_{t}}\right)=\sum\limits _{\left\{ m|a_{n,t+1}=a_{m,t}\right\} }\left(R_{m}\left(\mathbf{a_{\mathbf{t}+1}}\right)-R_{m}\left(\mathbf{a_{\mathbf{t}}}\right)\right)\label{eq:5-2}
\end{equation}
is the rate difference caused by the additional interference to players
in the channel that player $n$ is moving to, and
\begin{equation}
\varDelta_{3}\left(\mathbf{a_{t}}\right)=\sum\limits _{\left\{ m|a_{n,t}=a_{m,t}\right\} }\left(R_{m}\left(\mathbf{a_{\mathbf{t+1}}}\right)-R_{m}\left(\mathbf{a_{\mathbf{t}}}\right)\right)\label{eq:5-3}
\end{equation}
is the rate difference caused by the reduced interference to players
in the channel that player $n$ is moving from.

All three are random variables with a distribution that depends on
$\mathbf{a_{t}}$. Specifically, if player $n$ is the acting player
at time $t$ and $BR_{\varepsilon,n}\left(\mathbf{\mathbf{a_{t}}}\right)=\mathbf{\mathbf{a_{t}}}$
then $\varDelta_{1}\left(\mathbf{a_{t}}\right)=\varDelta_{2}\left(\mathbf{a_{t}}\right)=\varDelta_{3}\left(\mathbf{a_{t}}\right)=0$.

Our main result is the convergence of the approximate BR dynamics
to an $\varepsilon$-PNE such that each $\varepsilon$-PNE is asymptotically
$\varepsilon$-optimal in the max-min sense, and as well in the mean-rate.
This result is stated as follows
\begin{thm}[Main Result]
\label{k(a)} Assume that the players' locations are independently
and uniformly randomly generated in some closed set $\mathcal{D}\subseteq\mathbb{R}^{2}$
with area $\left|\mathcal{D}\right|=\frac{1}{\lambda}$ for some $\lambda>0$.
Also assume that the schedule $\left\{ \mathcal{N}_{t}\right\} $
is non-degenerate. If $P_{\textrm{max}}=P_{0}\left(\frac{\log N}{N}\right)^{\frac{\alpha}{2}}$
for some $P_{0}>0$ and $\frac{N}{K}=l_{N}\geq1$ satisfies $l_{N}=o\left(\left(\frac{N}{\log N}\right)^{\frac{\alpha}{\alpha+2}}\right)$
then, for any $\varepsilon>0$, the approximate BR dynamics converge
to an $\varepsilon$-PNE in a finite time with a probability that
approaches 1 as $N\rightarrow\infty.$ Denote by $\mathcal{P}_{\varepsilon}$
the random set of $\varepsilon$-PNE of the realization game. For
every $\delta>0$
\begin{equation}
\Pr\left(\underset{\begin{array}{c}
\mathbf{a}^{*}\in\mathcal{P}_{\varepsilon},\,n\in\mathcal{N}\end{array}}{\max}\left|u_{n}\left(\mathbf{a}^{*}\right)-\log_{2}\left(1+\frac{g_{n,d(n)}P_{n}}{N_{0}}\right)\right|\leq\varepsilon+\delta\right)\rightarrow1\label{eq:16-1}
\end{equation}
as $N\rightarrow\infty$.
\end{thm}
The proof of Theorem 1 is delayed for the end of Section IV. According
to this theorem, the $\varepsilon$-PNE of asymptotically almost all
the interference games exhibit vanishing interference for each player.
This fact on its own should not be surprising. There are $l_{N}\geq1$
players on average in each channel. If $l_{N}$ is not growing too
fast, or for simplicity is just a constant, it should be possible
to design an allocation where all these sharing players are geometrically
separated enough in $\mathcal{D}$. Due to the decreasing transmission
power required for each player to communicate with his destination,
this geometrical separation guarantees negligible interference for
large enough $N$. A small enough interference is a desired property
of any well-functioning network. The essence of our result is in the
fact that approximate BR, that constitute a very simple and distributed
algorithm, naturally tend to converge to this desired outcome. In
other words, the dynamics of the interference game in this scenario
can be thought of as a distributed and adaptive implementation of
frequency band allocation over space (see \cite{Ni2011}).

\section{Submartingality of the Sum-Rate Process }

We begin this section by proving that the process $X(\mathbf{a_{t}})=\sum_{n=1}^{N}R\left(I_{n}\left(\mathbf{a}_{t}\right)\right)$
is indeed a submartingale. More exactly, the sum-rate process is a
submartingale for asymptotically almost all the interference games,
so we want to show that
\[
\underset{N\rightarrow\infty}{\lim}\Pr\left(\forall t\,,E\left\{ X(\mathbf{a_{t+1}})-X(\mathbf{a_{t}})|\mathbf{a_{t}}\right\} \geq0\right)=1.
\]
Note that the above probability is with respect to the channel gains
that are generated at random at $t=0$, based on the locations of
the players, and remain constant for all $t\geq0$. The expectation
is over the randomness of the acting player and his actions. This
probability is our measure to ``count the number'' of interference
games where convergence of the approximate BR occurs (how large the
set of converging interference games is).

\begin{figure}[tbh]
~~~~~~~~~~~~~~~~~~~~~~~~~~~~~~~~~~~~~~~~~~~~~~~~\includegraphics[width=7cm,height=5cm]{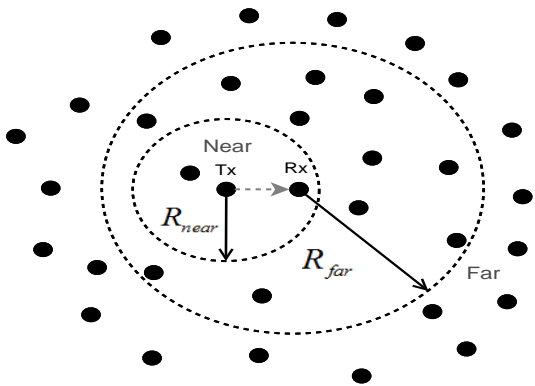}

\caption{\label{fig:NearFar}Illustration of the near and far regions.}
\end{figure}

The interference game characteristics are determined by the geometry
of its players. For instance, player $n$'s transmitter strongly interferes
players with a nearby destination, and player $n$'s destination undergoes
weak interference from far away transmitters. This is depicted in
Figure \ref{fig:NearFar}. We want the first set of near players to
be as small as possible while the second set of far players to be
as large as possible. We define these sets as follows
\begin{defn}
\label{NearFarSets}Assume the players are located in some closed
set $\mathcal{D}\subseteq\mathbb{R}^{2}$ with area $\left|\mathcal{D}\right|=\frac{1}{\lambda}$
for some $\lambda>0$. Define the set of near players, with respect
to player $n$'s transmitter, as
\begin{equation}
\mathcal{N}_{\textrm{near},n}=\Biggl\{ m\,|m\in\mathcal{N},\,r_{n,d\left(m\right)}\leq\frac{1}{\sqrt{\pi\lambda}}\left(\frac{\log N}{N}\right)^{\frac{\alpha}{2\alpha+4}},\left|\theta_{n,d\left(m\right)}\right|\leq\frac{\theta_{T}}{2}\,,d\left(m\right)\neq n\Biggr\}\label{eq:5-4}
\end{equation}
and the set of far players, with respect to player $n$'s receiver,
for some $q>1$, as
\begin{equation}
\mathcal{N}_{\textrm{far},n}=\mathcal{N}\setminus\left\{ m\,|m\in\mathcal{N},\,r_{m,d\left(n\right)}\leq\sqrt{\frac{1}{2\theta_{R}\lambda}\frac{q-1}{ql_{N}-1}},\thinspace\left|\theta_{d\left(n\right),m}\right|\leq\frac{\theta_{R}}{2}\right\} \label{eq:5-5}
\end{equation}
\end{defn}
The next lemma provides asymptotic probabilistic bounds for the cardinality
of these two sets.
\begin{lem}
\label{NearFar}Let $\frac{N}{K}=l_{N}\geq1$ satisfy $l_{N}=o\left(\left(\frac{N}{\log N}\right)^{\frac{\alpha}{\alpha+2}}\right)$.
Assume that the players' locations are independently and uniformly
randomly generated in some closed set $\mathcal{D}\subseteq\mathbb{R}^{2}$
with area $\left|\mathcal{D}\right|=\frac{1}{\lambda}$ for some $\lambda>0$.
Denote $N_{\textrm{near},n}=\left|\mathcal{N}_{\textrm{near},n}\right|$
and $N_{\textrm{far},n}=\left|\mathcal{N}_{\textrm{far},n}\right|$.
Then
\begin{equation}
\underset{N\rightarrow\infty}{\lim}\Pr\left(\underset{n\in\mathcal{N}}{\max}\,N_{\textrm{near},n}\leq\frac{\theta_{T}}{\pi}\left(N^{2}\log^{\alpha}N\right)^{\frac{1}{\alpha+2}}\right)=1\label{eq:6-2}
\end{equation}
and for any $q>1$
\begin{equation}
\underset{N\rightarrow\infty}{\lim}\Pr\left(\underset{n\in\mathcal{N}}{\min\,}N_{\textrm{far},n}\geq\left(1-\frac{1}{2}\frac{q-1}{ql_{N}-1}\right)N\right)=1\label{eq:7-2}
\end{equation}
where $\theta_{T}$ is the transmission beamforming angle (see Definition
\ref{Channel Gains}).
\end{lem}
\begin{IEEEproof}
Denote the sector originated at player $n$ with angle $\theta$ (around
an arbitrary line) and radius $t$ by $\Phi_{t,\theta,n}$, and the
number of players in it (not including player $n$) by $N_{t,\theta,n}$.
Due to the players' locations being uniformly and independently distributed,
for each $n$ and $m$ we have
\begin{equation}
\Pr\left(m\in\Phi_{t,\theta,n}\right)=\Pr\left(r_{n,m}\leq t\cdot I_{d}\left\{ \left|\theta_{n,m}\right|\leq\frac{\theta}{2}\right\} \right)\underset{(a)}{\leq}\frac{\left|\Phi_{t,\theta,n}\right|}{\left|\mathcal{D}\right|}\underset{(b)}{=}\frac{\lambda\theta t^{2}}{2}\label{eq:6-1}
\end{equation}
where (a) follows from the possibility that part of $\Phi_{t,\theta,n}$,
which has area $\frac{\theta t^{2}}{2}$, may fall outside $\mathcal{D}$
and (b) from $\left|\mathcal{D}\right|=\frac{1}{\lambda}$. From \eqref{eq:6-1}
we conclude that the random variable $N_{t,\theta,n}$ is stochastically
dominated by a binomially random variable with $N$ trials and success
probability $p=\frac{\lambda\theta t^{2}}{2}$, denoted $Y_{p}$.
This is due to the fact there are actually only $N-1$ trials, and
by \eqref{eq:6-1}, the success probability is smaller than $p$.
Using this fact we get from Theorem A.1.12 in \cite{Alon2015} that
for all $N>0$ and $\beta>1$
\begin{equation}
\Pr\left(N_{t,\theta,n}\geq\beta pN\right)\leq\Pr\left(Y_{p}\geq\beta pN\right)\leq\left(e^{\beta-1}\beta^{-\beta}\right)^{pN}.\label{eq:7-1}
\end{equation}
To prove \eqref{eq:6-2}, observe that $N_{\textrm{near},n}$ is distributed
like $N_{t,\theta,n}$ for $t=\frac{1}{\sqrt{\pi\lambda}}\left(\frac{\log N}{N}\right)^{\frac{\alpha}{2\alpha+4}}$
and $\theta=\theta_{T},$ so $p=\frac{\theta_{T}}{2\pi}\left(\frac{\log N}{N}\right)^{\frac{\alpha}{\alpha+2}}$.
By substituting $n=\frac{\theta_{T}}{\pi}\left(N^{2}\log^{\alpha}N\right)^{\frac{1}{\alpha+2}}=2pN$
in \eqref{eq:7-1} we obtain for all $N>0$
\begin{equation}
\Pr\left(N_{\textrm{\textrm{near}},1}\geq\frac{\theta_{T}}{\pi}\left(N^{2}\log^{\alpha}N\right)^{\frac{1}{\alpha+2}}\right)\leq\left(e\cdot2^{-2}\right)^{\frac{\theta_{T}}{2\pi}\left(N^{2}\log^{\alpha}N\right)^{\frac{1}{\alpha+2}}}.\label{eq:9}
\end{equation}
We conclude by the union bound that
\begin{multline}
\Pr\left(\underset{n\in\mathcal{N}}{\max\,}N_{\textrm{near},n}\geq\frac{\theta_{T}}{\pi}\left(N^{2}\log^{\alpha}N\right)^{\frac{1}{\alpha+2}}\right)=\Pr\left(\bigcup_{n=1}^{N}\left\{ N_{\textrm{near},n}\geq\frac{\theta_{T}}{\pi}\left(N^{2}\log^{\alpha}N\right)^{\frac{1}{\alpha+2}}\right\} \right)\leq\\
N\Pr\left(N_{\textrm{\textrm{near}},1}\geq\frac{\theta_{T}}{\pi}\left(N^{2}\log^{\alpha}N\right)^{\frac{1}{\alpha+2}}\right)\leq N\left(e\cdot2^{-2}\right)^{\frac{\theta_{T}}{2\pi}\left(N^{2}\log^{\alpha}N\right)^{\frac{1}{\alpha+2}}}\label{eq:10-2}
\end{multline}
so $\underset{N\rightarrow\infty}{\lim}\Pr\left(\underset{n\in\mathcal{N}}{\max}\,N_{\textrm{near},n}\leq\frac{\theta_{T}}{\pi}\left(N^{2}\log^{\alpha}N\right)^{\frac{1}{\alpha+2}}\right)=1$.

To prove \eqref{eq:7-2}, observe that we can write $N_{\textrm{far},n}=\left(N-1\right)-\tilde{N_{n}},$
where $\tilde{N_{n}}$ is distributed like $N_{t,\theta,d\left(n\right)}$
for $t=\sqrt{\frac{1}{2\theta_{R}\lambda}\frac{q-1}{ql_{N}-1}}$ and
$\theta=\theta_{R}$, so $p=\frac{1}{4}\frac{q-1}{ql_{N}-1}$. By
substituting $n=\frac{1}{2}\frac{q-1}{ql_{N}-1}N-1=\left(2-\frac{4}{N}\frac{ql_{N}-1}{q-1}\right)pN$
in \eqref{eq:7-1} and denoting $\eta=\frac{4}{N}\frac{ql_{N}-1}{q-1}$
we obtain that for all $N>0$
\begin{equation}
\Pr\left(\tilde{N}_{1}\geq\frac{1}{2}\frac{q-1}{ql_{N}-1}N-1\right)\leq\left(e^{1-\eta}\cdot\left(2-\eta\right)^{-\left(2-\eta\right)}\right)^{\frac{1}{4}\frac{q-1}{ql_{N}-1}N}\label{eq:11-1}
\end{equation}
We conclude by the union bound that
\begin{multline}
\Pr\left(\underset{n\in\mathcal{N}}{\min\,}N_{\textrm{far},n}\leq\left(1-\frac{1}{2}\frac{q-1}{ql_{N}-1}\right)N\right)=\Pr\left(\bigcup_{n=1}^{N}\left\{ N-N_{\textrm{far},n}\geq\frac{1}{2}\frac{q-1}{ql_{N}-1}N\right\} \right)\leq\\
N\Pr\left(\tilde{N}_{1}\geq\frac{1}{2}\frac{q-1}{ql_{N}-1}N-1\right)\leq N\left(e^{1-\eta}\cdot\left(2-\eta\right)^{-\left(2-\eta\right)}\right)^{\frac{1}{4}\frac{q-1}{ql_{N}-1}N}\label{eq:13-1}
\end{multline}
so $\underset{N\rightarrow\infty}{\lim}\Pr\left(\underset{n\in\mathcal{N}}{\min\,}N_{\textrm{far},n}\geq\left(1-\frac{1}{2}\frac{q-1}{ql_{N}-1}\right)N\right)=1$
(since $\eta\rightarrow0$ as $N\rightarrow\infty$).
\end{IEEEproof}
The following simple lemma will be useful for some of our calculations.
\begin{lem}
\label{loglog}For any $a,b,\bigtriangleup I>0$,
\begin{equation}
\log_{2}\left(1+\frac{a}{b}\right)-\log_{2}\left(1+\frac{a}{b+\bigtriangleup I}\right)\leq\frac{a}{a+b}\frac{\bigtriangleup I}{b\ln\left(2\right)}\label{eq:12}
\end{equation}
\end{lem}
\begin{IEEEproof}
This follows by
\begin{multline}
\log_{2}\left(1+\frac{a}{b}\right)-\log_{2}\left(1+\frac{a}{b+\bigtriangleup I}\right)=\log_{2}\left(1+\frac{\frac{a}{b}-\frac{a}{b+\bigtriangleup I}}{1+\frac{a}{b+\bigtriangleup I}}\right)=\log_{2}\left(1+\frac{a}{b}\frac{\bigtriangleup I}{a+\bigtriangleup I+b}\right)\leq\\
\log_{2}\left(1+\frac{a}{a+b}\frac{\bigtriangleup I}{b}\right)\underset{(a)}{\leq}\frac{a}{a+b}\frac{\bigtriangleup I}{b\ln\left(2\right)}\label{eq:10-1}
\end{multline}
where (a) is due to $\log_{2}\left(1+x\right)\leq\frac{x}{\ln\left(2\right)}$
for all $x\geq0$.
\end{IEEEproof}
Now we turn to exploit the above results for our purpose of showing
the submartingality of the sum-rate process. The following lemma shows
that the set $B_{\varepsilon}\left(\mathbf{a}_{-n}\right)$ is asymptotically
large, and each action in it is asymptotically approximately optimal
(interference free). This supplies the necessary averaging effect
over the random choice of an action among the approximately best actions.
\begin{lem}
\label{There are many good actions}Assume that the players' locations
are independently and uniformly randomly generated in some closed
set $\mathcal{D}\subseteq\mathbb{R}^{2}$ with area $\left|\mathcal{D}\right|=\frac{1}{\lambda}$
for some $\lambda>0$. If $P_{\textrm{max}}=P_{0}\left(\frac{\log N}{N}\right)^{\frac{\alpha}{2}}$
for some $P_{0}>0$ and $\frac{N}{K}=l_{N}\geq1$ satisfies $l_{N}=o\left(\left(\frac{N}{\log N}\right)^{\frac{\alpha}{\alpha+2}}\right)$,
then for any $q>1$ and $\varepsilon>0$
\begin{equation}
\underset{N\rightarrow\infty}{\lim}\Pr\left(\forall\mathbf{a},\,\underset{n\in\mathcal{N}}{\min}\left|B_{\varepsilon}\left(\mathbf{a}_{-n}\right)\right|\geq\frac{1}{2l_{N}}\left(1-\frac{1}{q}\right)N\right)=1.\label{eq:11-2}
\end{equation}
Furthermore,
\begin{equation}
\underset{\begin{array}{c}
n\in\mathcal{N}\,\mathbf{a\in}A_{1}\times...\times A_{N}\end{array}}{\max}\left|u_{n}\left(BR_{n}\left(\mathbf{a}\right)\right)-\log_{2}\left(1+\frac{g_{n,d(n)}P_{n}}{N_{0}}\right)\right|\rightarrow0\label{eq:13-3}
\end{equation}
in probability as $N\rightarrow\infty$.
\end{lem}
\begin{IEEEproof}
See Appendix A.
\end{IEEEproof}
No doubt that a deviating player can only improve his utility, but
the following lemma shows that on average over his chosen action,
he also causes asymptotically negligible interference to other players.
This establishes that the sum-rate process is indeed a submartingale.
\begin{lem}
\label{Steps are good in average}Assume that the players' locations
are independently and uniformly randomly generated in some closed
set $\mathcal{D}\subseteq\mathbb{R}^{2}$ with area $\left|\mathcal{D}\right|=\frac{1}{\lambda}$
for some $\lambda>0$. Also assume that $P_{\textrm{max}}=P_{0}\left(\frac{\log N}{N}\right)^{\frac{\alpha}{2}}$
for some $P_{0}>0$ and that $\frac{N}{K}=l_{N}\geq1$ satisfies $l_{N}=o\left(\left(\frac{N}{\log N}\right)^{\frac{\alpha}{\alpha+2}}\right)$.
Let $\varepsilon>0$ and the acting player at time $t$ be player
$n$. Denote by $E_{n}$ the expectation over his chosen channel.
Then the probability that for every $\mathbf{a}_{t}$ and $n$ such
that $\mathbf{a_{\mathbf{t}+1}}=BR_{\varepsilon,n}\left(\mathbf{a}_{t}\right)\neq\mathbf{a}_{t}$,
for any $t\geq0$
\begin{equation}
E_{n}\left\{ \varDelta_{1}\left(\mathbf{a_{t}}\right)+\varDelta_{2}\left(\mathbf{a_{t}}\right)|\,\mathbf{a_{\mathbf{t}}}\right\} >0\label{eq:11}
\end{equation}
approaches 1 a $N\rightarrow\infty$, where the probability is with
respect to the random interference game, and $\varDelta_{1}\left(\mathbf{a_{t}}\right),\varDelta_{2}\left(\mathbf{a_{t}}\right)$
are defined in \eqref{eq:5-1}, \eqref{eq:5-2}.
\end{lem}
\begin{IEEEproof}
See Appendix B.
\end{IEEEproof}
We conclude this section by proving our main result.
\begin{IEEEproof}[Proof of Theorem 1]
Let $\varepsilon>0$. According to Lemma \ref{Steps are good in average}
(see \eqref{eq:15-1}) the probability that for every $\mathbf{a}_{t}$
and $n$ such that $\mathbf{a_{\mathbf{t}+1}}=BR_{\varepsilon,n}\left(\mathbf{a}_{\mathbf{t}}\right)\neq\mathbf{a}_{t}$,
the following holds for all $t\geq0$
\begin{equation}
E_{n}\left\{ \varDelta_{1}\left(\mathbf{a_{t}}\right)+\varDelta_{2}\left(\mathbf{a_{t}}\right)+\varDelta_{3}\left(\mathbf{a_{t}}\right)|\,\mathbf{a_{\mathbf{t}}}\right\} \geq\frac{\varepsilon}{4}\label{eq:19-1}
\end{equation}
approaches 1 as $N\rightarrow\infty$, because $\varDelta_{3}\left(\mathbf{a_{t}}\right)\geq0$
for each $\mathbf{a}_{t}$. Clearly $X\left(\mathbf{a_{\mathbf{t}+1}}\right)=X\left(\mathbf{a_{\mathbf{t}}}\right)$
if $\mathbf{a_{\mathbf{t}}}$ is an $\varepsilon$-PNE. Summing over
all $n\in\mathcal{N}_{t}$ and dividing by $\frac{1}{N_{t}}$ we obtain
that
\begin{equation}
\underset{N\rightarrow\infty}{\lim}\Pr\left(\forall t\,,E\left\{ X(\mathbf{a_{t+1}})-X(\mathbf{a_{t}})|\mathbf{a_{t}}\right\} \geq0\right)=1.\label{eq:20-1}
\end{equation}
Observe that the sum-rate process is bounded, since for each $t\geq0$,
with probability 1
\begin{equation}
\left|X\left(\mathbf{a_{\mathbf{t}}}\right)\right|\leq\sum_{n=1}^{N}\log_{2}\left(1+\frac{g_{n,d\left(n\right)}P_{n}}{N_{0}}\right).\label{eq:28-5}
\end{equation}
If indeed \eqref{eq:19-1} holds, then from the Martingale Convergence
Theorem \cite[Page 89, Theorem 5.14]{Breiman1992} we conclude that
there exists a random variable $X$ such that $X\left(\mathbf{a_{\mathbf{t}}}\right)\rightarrow X$
almost surely and $E\left\{ \left|X\right|\right\} <\infty$. Due
to \eqref{eq:28-5}, the almost sure convergence also implies $L^{1}$
convergence (bounded random variables are trivially uniformly integrable),
so for $\varepsilon'=\frac{\varepsilon}{8N^{2}}$ there exists an
$T>0$ such that for each $t\geq T$
\begin{equation}
E\left\{ \left|X\left(\mathbf{a}_{t+1}\right)-X\left(\mathbf{a}_{t}\right)\right|\right\} \underset{(a)}{\leq}E\left\{ \left|X\left(\mathbf{a}_{t+1}\right)-X\right|\right\} +E\left\{ \left|X\left(\mathbf{a}_{t}\right)-X\right|\right\} <2\varepsilon'\label{eq:26}
\end{equation}
where (a) follows from the triangle inequality. Denote by $\mathcal{P}_{\varepsilon}$
the set of $\varepsilon$-PNE points of the game. We conclude that
with probability (with respect to the random game) that approaches
1 as $N\rightarrow\infty$
\begin{multline}
E\left\{ X\left(\mathbf{a_{\mathbf{T}+1}}\right)-X\left(\mathbf{a_{\mathbf{T}}}\right)\right\} \underset{(a)}{=}\sum_{\mathbf{a}\in A_{1}\times...\times A_{n}}E\left\{ X\left(\mathbf{a_{\mathbf{T}+1}}\right)-X\left(\mathbf{a_{\mathbf{T}}}\right)\,|\,\mathbf{a_{\mathbf{T}}}=\mathbf{a}\right\} \Pr\left(\mathbf{a_{\mathbf{T}}=\mathbf{a}}\right)=\\
\sum_{\mathbf{a\in}\mathcal{P}_{\varepsilon}}E\left\{ X\left(\mathbf{a_{\mathbf{T}+1}}\right)-X\left(\mathbf{a_{\mathbf{T}}}\right)\,|\,\mathbf{a_{\mathbf{T}}}=\mathbf{a}\right\} \Pr\left(\mathbf{a_{\mathbf{T}}=\mathbf{a}}\right)+\sum_{\mathbf{a\notin}\mathcal{P}_{\varepsilon}}E\left\{ X\left(\mathbf{a_{\mathbf{T}+1}}\right)-X\left(\mathbf{a_{\mathbf{T}}}\right)\,|\,\mathbf{a_{\mathbf{T}}}=\mathbf{a}\right\} \Pr\left(\mathbf{a_{\mathbf{T}}=\mathbf{a}}\right)\underset{(b)}{\geq}\\
\frac{\varepsilon}{4N_{t}}\sum_{\mathbf{a\notin}\mathcal{P}_{\varepsilon}}\Pr\left(\mathbf{a_{\mathbf{T}}=\mathbf{a}}\right)\geq\frac{\varepsilon}{4N}\Pr\left(\mathbf{a_{\mathbf{T}}\notin}\mathcal{P}_{\varepsilon}\right)\label{eq:27}
\end{multline}
where (a) follows from the law of total expectation and (b) is from
averaging \eqref{eq:19-1} over $N_{t}$, relying on the non-degeneracy
of the schedule $\left\{ \mathcal{N}_{t}\right\} $. Inequality (b)
is also due to $X\left(\mathbf{a_{\mathbf{T}+1}}\right)=X\left(\mathbf{a_{\mathbf{T}}}\right)$
with probability 1 for all $\mathbf{a_{\mathbf{T}}\in}\mathcal{P}_{\varepsilon}$.
Note that $\Pr\left(\mathbf{a_{\mathbf{T}}\notin}\mathcal{P}_{\varepsilon}\right)$
is with respect to the random player and his actions (and not the
random game). From \eqref{eq:27}
\begin{equation}
\Pr\left(\mathbf{a_{\mathbf{T}}\notin}\mathcal{P}_{\varepsilon}\right)\leq\frac{4N}{\varepsilon}E\left\{ X\left(\mathbf{a_{\mathbf{T}+1}}\right)-X\left(\mathbf{a_{\mathbf{T}}}\right)\right\} \leq\frac{4N}{\varepsilon}E\left\{ \left|X\left(\mathbf{a}_{T+1}\right)-X\left(\mathbf{a}_{T}\right)\right|\right\} \underset{(a)}{<}\frac{1}{N}\label{eq:24-1}
\end{equation}
where (a) is from \eqref{eq:26}, and hence $\Pr\left(\mathbf{a_{\mathbf{T}}\notin}\mathcal{P}_{\varepsilon}\right)\rightarrow0$
as $N\rightarrow\infty$.

Part 2 follows by the definition of an $\varepsilon$-PNE, since for
each $n$ and each $\mathbf{a}^{*}\in\mathcal{P}_{\varepsilon}$
\begin{equation}
u_{n}\left(\mathbf{a}^{*}\right)\geq u_{n}\left(BR_{n}\left(\mathbf{a}^{*}\right)\right)-\varepsilon\label{eq:24}
\end{equation}
combining this with \eqref{eq:13-3} from Lemma \ref{There are many good actions},
we obtain \eqref{eq:16-1}. Finally, if $\underset{n\in\mathcal{N}}{\max}\left|u_{n}\left(\mathbf{a}^{*}\right)-\log_{2}\left(1+\frac{g_{n,d\left(n\right)}P_{n}}{N_{0}}\right)\right|\leq\varepsilon+\delta$
then also
\begin{equation}
\frac{1}{N}\sum_{n=1}^{N}u_{n}\left(\mathbf{a}^{*}\right)\geq\frac{1}{N}\sum_{n=1}^{N}\log_{2}\left(1+\frac{g_{n,d\left(n\right)}P_{n}}{N_{0}}\right)-\varepsilon-\delta.\label{eq:25}
\end{equation}
\end{IEEEproof}
It is worth mentioning that although the above theorem uses the non-degeneracy
of the schedule $\left\{ \mathcal{N}_{t}\right\} $, it is not necessary.
Any schedule that ensures that each player gets an opportunity to
act infinitely often is enough. This of course might be true even
if on some turns all the possible acting players do not want to change
their actions. Nevertheless, different schedules will result in different
convergence times in terms of steps. The conversion between the number
of steps until convergence to absolute time is dependent on the random
access scheme.

\section{Convergence Time }

Our results from last section regarding the average change in the
sum-rate in each step can be used to draw conclusions about the convergence
time of the dynamics. In (approximate) BR dynamics, each player only
considers the current strategy profile for his decision, and therefore
these dynamics induce a Markov chain on $\left\{ \mathbf{a_{\mathbf{t}}}\right\} $.
From this perspective, our result from the last section establishes
that this Markov chain is absorbing (from every state it is possible
to reach an absorbing state), and hence converges to one of its absorbing
states ($\varepsilon$-PNE points). Note that this is equivalent to
the definition of a weakly acyclic game. This Markovity is the key
to the following lemma.
\begin{lem}
\label{Iterating on E}Let $\left\{ X(\mathbf{a_{t}})\right\} $ be
the sum-rate process induced by the approximate BR dynamics and $k\left(\mathbf{a}\right):A_{1}\times...\times A_{n}\rightarrow\mathbb{R}$.
If for each $\mathbf{a}$ and each $t\geq0$
\begin{equation}
E\left\{ X(\mathbf{a_{t+1}})-X(\mathbf{a_{t}})|\mathbf{a_{t}}=\mathbf{a}\right\} \geq k\left(\mathbf{a}\right)\label{eq:26-1}
\end{equation}
 then
\begin{equation}
E\left\{ X\left(\mathbf{a_{T}}\right)\thinspace|\,\mathbf{a}_{0}\right\} \geq X\left(\mathbf{a}_{0}\right)+E\left\{ \sum_{t=0}^{T-1}k\left(\mathbf{a}_{t}\right)|\,\,\mathbf{a}_{0}\right\} \label{eq:27-1}
\end{equation}
\end{lem}
\begin{IEEEproof}
We can iterate on \eqref{eq:26-1} in the following manner
\begin{multline}
E\left\{ X(\mathbf{a_{t+2}})-X(\mathbf{a_{t}})|\mathbf{a_{t}},...,\mathbf{a}_{0}\right\} =E\left\{ X(\mathbf{a_{t+2}})-X(\mathbf{a_{t+1}})|\mathbf{a_{t}},...,\mathbf{a}_{0}\right\} +E\left\{ X\left(\mathbf{a_{t+1}}\right)-X\left(\mathbf{a_{t}}\right)|\mathbf{a_{t}},...,\mathbf{a}_{0}\right\} \underset{(a)}{=}\\
E\left\{ E\left\{ X\left(\mathbf{a_{t+2}}\right)-X(\mathbf{a_{t+1}})|\mathbf{a_{t+1}},...,\mathbf{a}_{0}\right\} |\mathbf{a_{t}},...,\mathbf{a}_{0}\right\} +E\left\{ X\left(\mathbf{a_{t+1}}\right)-X\left(\mathbf{a_{t}}\right)|\mathbf{a_{t}},...,\mathbf{a}_{0}\right\} \underset{(b)}{=}\\
E\left\{ E\left\{ X\left(\mathbf{a_{t+2}}\right)-X(\mathbf{a_{t+1}})|\mathbf{a_{t+1}}\right\} |\mathbf{a_{t}}\right\} +E\left\{ X\left(\mathbf{a_{t+1}}\right)-X\left(\mathbf{a_{t}}\right)|\mathbf{a_{t}}\right\} \underset{(c)}{\geq}E\left\{ k\left(\mathbf{a_{t+1}}\right)+k\left(\mathbf{a_{t}}\right)|\mathbf{a_{t}}\right\} \label{eq:28}
\end{multline}
where (a) follows from the generalized law of total expectation, (b)
from the Markovity of the process and (c) from \eqref{eq:26-1}. Inequality
\eqref{eq:27-1} follows by iterating $T$ times on \eqref{eq:26-1}
starting from $t=0$.
\end{IEEEproof}
Using the above lemma we prove the next proposition.
\begin{prop}
\label{WorstCaseConvergence}Assume that the players' locations are
independently and uniformly randomly generated in some closed set
$\mathcal{D}\subseteq\mathbb{R}^{2}$ with area $\left|\mathcal{D}\right|=\frac{1}{\lambda}$
for some $\lambda>0$. Also assume that $P_{\textrm{max}}=P_{0}\left(\frac{\log N}{N}\right)^{\frac{\alpha}{2}}$
for some $P_{0}>0$ and that $\frac{N}{K}=l_{N}\geq1$ satisfies $l_{N}=o\left(\left(\frac{N}{\log N}\right)^{\frac{\alpha}{\alpha+2}}\right)$.
Let $\varepsilon>0$. Denote by $\mathcal{P}_{\varepsilon}$ the set
of $\varepsilon$-PNE points of the game. Denote $\overline{C}=\frac{1}{N}\sum_{n=1}^{N}\log_{2}\left(1+\frac{g_{n,d\left(n\right)}P_{n}}{N_{0}}\right)$.
Define the convergence time of the approximate BR dynamics for an
interference game by $T_{\textrm{con}}=\underset{t\geq0}{\min}\left\{ t\,|\,\mathbf{a_{t}\in\mathcal{P}_{\varepsilon}}\right\} $.
Assume $T_{\textrm{con}}<\infty$ (which happens with probability
that approaches 1 as $N\rightarrow\infty$, with respect to the random
game).

\begin{enumerate}
\item If $\left\{ \mathcal{N}_{t}\right\} $ is non-degenerate then $\Pr\left(T_{\textrm{con}}\geq L\frac{4\overline{C}}{\varepsilon}N^{2}\right)\leq\frac{1}{L}$
for each $L>0$.
\item If $\mathcal{N}_{t}=\left\{ n\,|\,BR_{\varepsilon,n}\left(\mathbf{a}_{\mathbf{t}}\right)\neq\mathbf{a}_{t}\right\} $
for each $t\geq0$ then $\Pr\left(T_{\textrm{con}}\geq L\frac{4\overline{C}}{\varepsilon}N\right)\leq\frac{1}{L}$
for each $L>0$.
\end{enumerate}
where the probabilities are with respect to the random players and
actions.
\end{prop}
\begin{IEEEproof}
Following the proof of Theorem \ref{k(a)} (see \eqref{eq:19-1}),
we know that the probability that for every $\mathbf{a}_{t}$ and
$n$ such that $\mathbf{a_{\mathbf{t}+1}}=BR_{\varepsilon,n}\left(\mathbf{a}_{\mathbf{t}}\right)\neq\mathbf{a}_{t}$,
the following holds for all $t\geq0$
\begin{equation}
E_{n}\left\{ X\left(\mathbf{a_{t+1}}\right)-X\left(\mathbf{a_{t}}\right)|\mathbf{a_{t}}\right\} \geq\frac{\varepsilon}{4}\label{eq:29-1}
\end{equation}
approaches 1 as $N\rightarrow\infty$, where $E_{n}$ denotes the
expectation over the random action of player $n$. Define the set
of deviating players at time $t$ as $\mathcal{N}_{d,t}=\left\{ n\,|\,n\in\mathcal{N}_{t},\,BR_{\varepsilon,n}\left(\mathbf{a}_{t}\right)\neq\mathbf{a}_{t}\right\} $
and denote their number by $N_{d,t}$. From the non-degeneracy of
$\left\{ \mathcal{N}_{t}\right\} $ we know that $N_{d,t}\geq1$ and
hence, by averaging over $\mathcal{N}_{t}$ and using \eqref{eq:29-1},
we obtain for each $t<T_{\textrm{con}}$
\begin{equation}
E\left\{ X\left(\mathbf{a_{t+1}}\right)-X\left(\mathbf{a_{t}}\right)|\mathbf{a_{t}}\right\} \geq\frac{N_{d,t}}{4N_{t}}\varepsilon.\label{eq:29-2}
\end{equation}
Therefore by Lemma \ref{Iterating on E} we get, for each $\mathbf{a}_{0}\in A_{1}\times...\times A_{n}$
\begin{equation}
E\left\{ X\left(\mathbf{a}_{T_{\textrm{con}}}\right)\thinspace|\,\mathbf{a}_{0}\right\} \geq X\left(\mathbf{a}_{0}\right)+E\left\{ \sum_{t=0}^{T_{\textrm{con}}-1}k\left(\mathbf{a}_{t}\right)|\,\mathbf{a}_{0}\right\} \underset{(a)}{\geq}\frac{\varepsilon}{4}E\left\{ \sum_{t=0}^{T_{\textrm{con}}-1}\frac{N_{d,t}}{N_{t}}\,|\mathbf{a}_{0}\right\} \geq\frac{\varepsilon}{4}\underset{t}{\min}\frac{N_{d,t}}{N_{t}}E\left\{ T_{\textrm{con}}|\,\mathbf{a}_{0}\right\} \label{eq:29-3}
\end{equation}
where (a) follows from $X\left(\mathbf{a}_{0}\right)\geq0$ and $k\left(\mathbf{a}_{t}\right)=\frac{\varepsilon}{4}\frac{N_{d,t}}{N_{t}}$.
Now define
\begin{equation}
\hat{T}=\frac{4N\overline{C}}{\varepsilon}\frac{1}{\underset{t}{\min}\frac{N_{d,t}}{N_{t}}}\label{eq:29-5}
\end{equation}
and observe that we must have $E\left\{ T_{\textrm{con}}|\,\mathbf{a}_{0}\right\} \leq\hat{T}$
for each $\mathbf{a}_{0}\in A_{1}\times...\times A_{n}$, otherwise
by \eqref{eq:29-3} we get $E\left\{ X\left(\mathbf{a_{T_{\textrm{con}}}}\right)\thinspace|\,\mathbf{a}_{0}\right\} >N\overline{C}$,
which is a contradiction (see \eqref{eq:28-5}). By Markov inequality
we get, for each $L>0$
\begin{equation}
\Pr\left(T_{\textrm{con}}\geq L\hat{T}\right)\leq\frac{E\left\{ T_{\textrm{con}}\right\} }{L\hat{T}}\leq\frac{1}{L}.\label{eq:30-1}
\end{equation}
The two parts of this proposition follow from \eqref{eq:29-5}, because
if $\left\{ \mathcal{N}_{t}\right\} $ is non-degenerate then $\underset{t}{\min}\frac{N_{d,t}}{N_{t}}\geq\frac{1}{N}$,
and if $\mathcal{N}_{t}=\left\{ n\,|\,BR_{\varepsilon,n}\left(\mathbf{a}_{t}\right)\neq\mathbf{a}_{t}\right\} $
for each $t\geq0$ then $\underset{t}{\min}\frac{N_{d,t}}{N_{t}}=1$.
\end{IEEEproof}

\section{Simulation Results}

We simulated $K$ channels and $N$ players independently and uniformly
at random on a two-dimensional disk with radius $R_{disk}=10m$. The
channel gains were chosen according to \eqref{eq:1-1} with wavelength
$\nu=\frac{c}{2.4\cdot10^{9}}$ and $\alpha=3.5$. Each player's destination
was drawn randomly from his five nearest neighbors. The transmission
power of each player was chosen such that $SNR=20dB$ in the absence
of interference. These two features are essentially a power control
mechanism that makes $P_{n}$ proportional to $\frac{1}{N^{\frac{\alpha}{2}}}$
on average over the random network, as it appears in our system model.
In each iteration we chose a player at random from the set of players
who want to switch a channel, i.e., $\mathcal{N}_{t}=\left\{ n\,|\,BR_{\varepsilon,n}\left(\mathbf{a}_{\mathbf{t}}\right)\neq\mathbf{a}_{t}\right\} $.
This player performed his approximate BR with some $\varepsilon$.
Unless otherwise stated, the beamforming angles were $\theta_{T}=\theta_{R}=2\pi$.

\begin{figure}[t]
~~~~~~~~~~~~~~~~~~~~~~~~~~~~~~~~~~~~~\includegraphics[width=9cm,height=6cm]{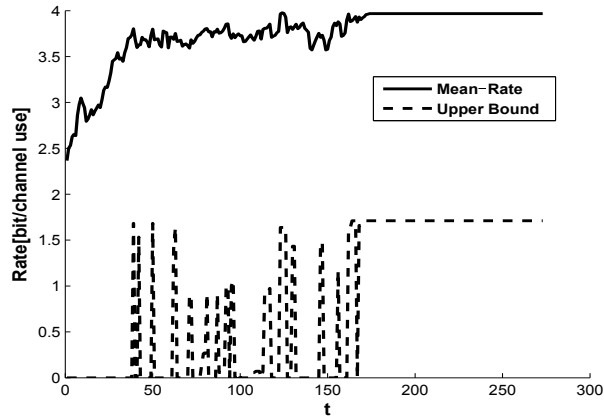}

\caption{\label{fig:single}The mean and minimal rate of a single realization.}
\end{figure}

Figure \ref{fig:single} shows the convergence for a single realization
with $N=50$, $K=5$ and $\varepsilon=0.1$. The beamforming angles
were chosen to be $\theta_{T}=\frac{2\pi}{3}$ and $\theta_{R}=2\pi$.
The martingale property of the mean-rate is clearly shown, as some
of the iterations result in a mean-rate decrease. We can see that
although $N$ is relatively small, the average mean-rate at equilibrium
is not significantly smaller than $\log_{2}\left(1+100\right)=6.66$
and that the minimal rate converges to some reasonable value.

\begin{figure}[t]
~~~~~~~~~~~~~~~~~~~~~~~~~~~~~~~~~~~~~\includegraphics[width=9cm,height=6cm]{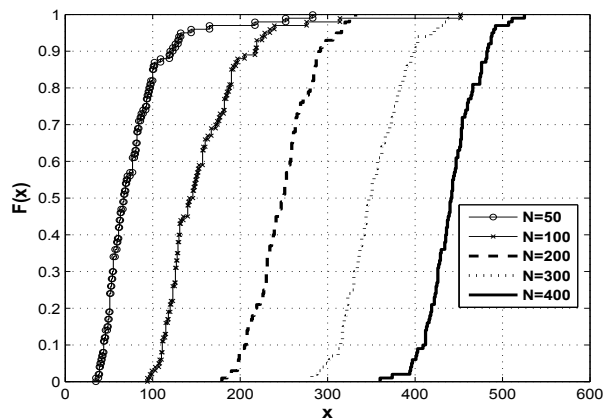}

\caption{\label{fig:VersusN}Empirical CDFs of the convergence time for different
$N$, with $l_{N}=10$.}
\end{figure}

Figure \ref{fig:VersusN} shows the empirical CDFs, based on 100 interference
game realizations, of the convergence time for $N=50,100,200,300,400$,
$\varepsilon=0.1$ and $l_{N}=\frac{N}{K}=10$. It is evident that
convergence to an $\varepsilon-$PNE indeed occurs, and it only requires
a few actions on average from each player. The average convergence
times are $78,156,249,353,443$, which are better than linear with
$N$ as expected by Proposition \ref{WorstCaseConvergence}.

\begin{figure}[t]
~~~~~~~~~~~~~~~~~~~~~~~~~~~~~~~~~~~~~\includegraphics[width=9cm,height=6cm]{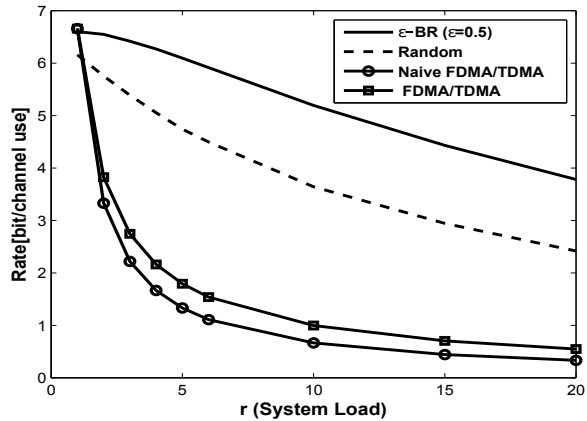}

\caption{\label{fig:Versusr}Mean rate for different $l_{N}$, with $N=300$.}
\end{figure}

Figure \ref{fig:Versusr} shows the mean rate as a function of the
system load $l_{N}$, based on 100 interference game realizations,
for $N=300$ , $\varepsilon=0.5$ and $l_{N}=1,2,3,4,5,6,10,15,20$.
We compare our results to those of a random allocation and a time-frequency
multiple access allocation (TDMA/FDMA or orthogonal FDMA). Trivially,
a TDMA/FDMA scheme achieves $R\left(l_{N}\right)=\frac{\log_{2}\left(101\right)}{l_{N}}$
for all players or $R\left(l_{N}\right)=\frac{\log_{2}\left(1+l_{N}\cdot100\right)}{l_{N}}$
if the power limitation is over a frame. Our approximate BR outperforms
all of these schemes for any $l_{N}$, and with an increasing ratio.
For example, for $l_{N}=2$ our mean-rate is 1.7 times better than
that of TDMA/FDMA and for $l_{N}=20$ the ratio is 6.9.

\begin{figure}[t]
~~~~~~~~~~~~~~~~~~~~~~~~~~~~~~~~~~~~~\includegraphics[width=9cm,height=6cm]{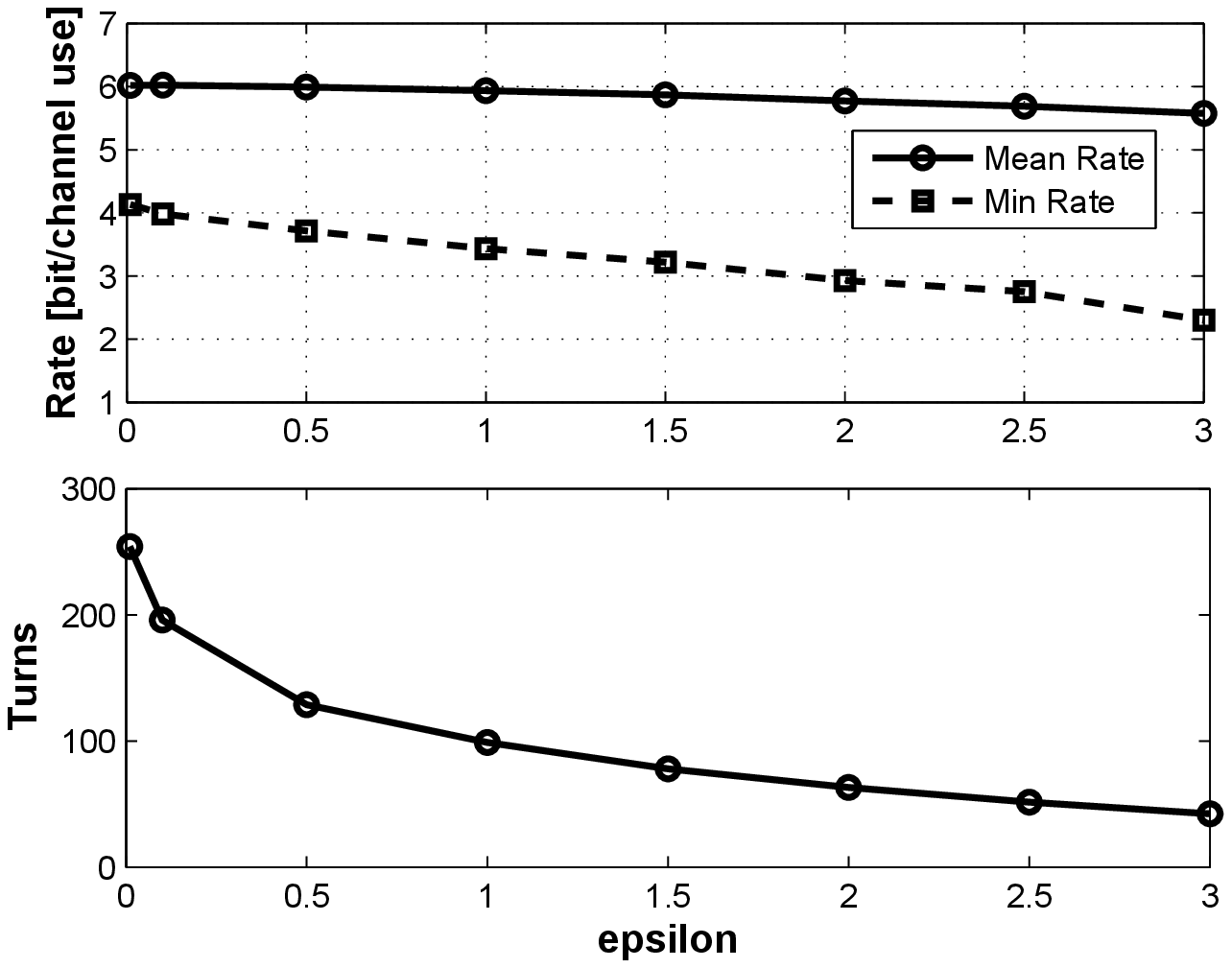}

\caption{\label{fig:Things_Versus_Epsilon}Rates and convergence time for different
$\varepsilon$.}
\end{figure}

Figure \ref{fig:Things_Versus_Epsilon} presents the trade-off introduced
by $\varepsilon$ between the convergence time and the rates. We used
$N=200$ and $K=50$, for $\varepsilon=0.01,0.1,0.5,1,1.5,2,2.5$
and averaged the results over 200 realizations. Clearly, the convergence
time decreases (improves) with $\varepsilon$ and so do the rates
(get worse). For both the convergence time and the rates the effect
of $\varepsilon$ seems to be slower than linear with a unit slope.
This implies that the aforementioned trade-off is non-trivial. While
the rate improvement of the acting player is always at least $\frac{\varepsilon}{2}$
, in many of the turns (especially the initial ones) it is much larger
and unaffected by $\varepsilon$. Regarding the rates, the mean rates
are much less affected by $\varepsilon$ than the minimal rates. This
is to be expected since the distribution of the rates of the channels
in $B_{\varepsilon}\left(\mathbf{a}_{-n}\right)$ is far from being
uniform between $R_{n}\left(0\right)-\varepsilon$ and $R_{n}\left(0\right)$.
Most of the channels in $B_{\varepsilon}\left(\mathbf{a}_{-n}\right)$
result in a rate that is close to $R_{n}\left(0\right)$, with nothing
to do with $\varepsilon$. This determines the average rate of the
players. On the other hand, the worst channels in $B_{\varepsilon}\left(\mathbf{a}_{-n}\right)$,
that result in the minimal rates, are (asymptotically) bounded from
below by $R_{n}\left(0\right)-\varepsilon$, and were originally chosen
by the corresponding players to be better than $R_{n}\left(0\right)-\frac{\varepsilon}{2}$.
Indeed, we see that the minimal rates decrease with $\varepsilon$
with a slope between $\frac{1}{2}$ and $1$.


\section{Conclusion}

In this work, we proved the convergence of asynchronous approximate
BR dynamics in a class of games called interference games, which model
a natural conflict in many wireless network scenarios. We generated
an interference game at random by generating the players' locations
uniformly and independently at random at some area $\mathcal{D}$,
and set the channel gains accordingly. Although interference games
tend not to be potential games, we were able to show that the approximate
BR dynamics converge to an approximate Nash equilibrium (NE) for almost
all of the interference games, asymptotically in the number of players.
The sum-rate of the resulting approximate NE is asymptotically approximately
close to the optimal, and so is the minimal rate. This stems from
the submartingality of the sum-rate process that our dynamics induce
on the strategy profiles. This approach also enables a simple analysis
of the convergence time. Simulations support our results and show
convergence to an approximate PNE that only requires a small number
of updates from each player. This also implies that our dynamics are
inherently robust to the addition of new players. If, at some point
in time, a new player arrives and chooses a channel, it likely requires
a small portion of players to change their channel until an $\varepsilon$-PNE
is maintained.

Our solution achieves self-configuration and low energy consumption,
both of which are desirable for small, simple devices that will take
part in the IoT and futuristic WSAN. Furthermore, such devices will
require simple protocols, so the distributed fashion of our resource
allocation scheme is of crucial value.

Last but not least, using probabilistic tools, by imposing a probability
space over the set of games, is a strong tool to analyze the dynamics
of a large class of games. This is a novel approach which can have
many other new applications.

\appendices{}

\section{Proof of Lemma \ref{There are many good actions}}
\begin{IEEEproof}
Let player $n$ be the current acting player and denote $N_{\textrm{far},n}=\left|\mathcal{N}_{\textrm{far},n}\right|$
(see Definition \ref{NearFarSets}). For some $q>1$, define the set
\begin{equation}
\mathcal{K}_{g,n}\left(\mathbf{a}_{-n}\right)=\left\{ k\,|\,a_{m}\neq k\,,\forall m\notin\mathcal{N}_{\textrm{far},n}\thinspace,N_{k}\left(\mathbf{a}_{-n}\right)<\left\lceil ql_{N}\right\rceil \right\} \label{eq:6}
\end{equation}
where $N_{k}\left(\mathbf{a}_{-n}\right)$ is the number of players
(apart from player $n$) in channel $k$. This is the set of channels
that contain only far players (from the receiver) and no more than
$\left\lceil ql_{N}\right\rceil -1$ players (apart from $n$). We
want to lower bound the cardinality of this set for all strategy profiles.
Players that are not far can occupy at most $N-N_{\textrm{far},n}$
channels, so each of them can reduce $\left|\mathcal{K}_{g,n}\left(\mathbf{a}_{-n}\right)\right|$
by at most one. Now we are left with $N_{\textrm{far},n}$ far players
and we need at least $\left\lceil ql_{N}\right\rceil $ of them to
reduce $\left|\mathcal{K}_{g,n}\left(\mathbf{a}_{-n}\right)\right|$
by one, so
\begin{equation}
\left|\mathcal{K}_{g,n}\left(\mathbf{a}_{-n}\right)\right|\geq K-\left(N-N_{\textrm{far},n}\right)-\left\lfloor \frac{N_{\textrm{far},n}}{\left\lceil ql_{N}\right\rceil }\right\rfloor \underset{(a)}{\geq}K+N_{\textrm{far},n}\left(1-\frac{1}{ql_{N}}\right)-N\label{eq:7}
\end{equation}
where in (a) we used the trivial inequality $\left\lfloor \frac{N_{\textrm{far},n}}{\left\lceil ql_{N}\right\rceil }\right\rfloor \leq\frac{N_{\textrm{far},n}}{ql_{N}}$.
By Lemma \ref{NearFar}
\begin{equation}
\underset{N\rightarrow\infty}{\lim}\Pr\Biggl(K+\left(1-\frac{1}{ql_{N}}\right)\underset{n\in\mathcal{N}}{\min\,}N_{\textrm{far},n}-N\geq\frac{1}{2l_{N}}\left(1-\frac{1}{q}\right)N\Biggr)=\underset{N\rightarrow\infty}{\lim}\Pr\left(\underset{n\in\mathcal{N}}{\min\,}N_{\textrm{far},n}\geq\left(1-\frac{1}{2}\frac{q-1}{ql_{N}-1}\right)N\right)=1\label{eq:8-1}
\end{equation}
In words, the probability that for each $n$ and $\mathbf{a}$ there
exist at least $\frac{1}{2l_{N}}\left(1-\frac{1}{q}\right)N$ channels
in $\mathcal{K}_{g,n}\left(\mathbf{a}_{-n}\right)$ goes to 1 as $N\rightarrow\infty$.
Now observe that for each $n$, each channel $k\in\mathcal{K}_{g,n}\left(\mathbf{a}_{-n}\right)$
satisfies
\begin{equation}
I_{n}\left(k,\mathbf{a}_{-n}\right)=\sum\limits _{\left\{ m|a_{m}=k\right\} }g_{m,d\left(n\right)}P_{m}\underset{(a)}{\leq}\sum\limits _{\left\{ m|a_{m}=k\right\} }\frac{GP_{m}}{r_{m,d\left(n\right)}^{\alpha}}\underset{(b)}{\leq}\frac{\lambda^{\frac{\alpha}{2}}GqP_{0}}{\left(\frac{1}{2\theta_{R}}\frac{q-1}{ql_{N}-1}\right)^{\alpha/2}}l_{N}\left(\frac{\log N}{N}\right)^{\frac{\alpha}{2}}\label{eq:9-1}
\end{equation}
where (a) and (b) are for the worst case where all the involved indicator
functions (see \eqref{eq:1-1}) are equal to 1 and $r_{m,d\left(n\right)}\geq\sqrt{\frac{1}{2\theta_{R}\lambda}\frac{q-1}{ql_{N}-1}}$
(otherwise $g_{m,d\left(n\right)}=0$). Inequality (b) also uses $P_{m}\leq P_{0}\left(\frac{\log N}{N}\right)^{\frac{\alpha}{2}}$
and the fact that there are no more than $\left\lceil ql_{N}\right\rceil -1\leq ql_{N}$
players in each of these channels. Denote $\gamma=\underset{n\in\mathcal{N}}{\max}\frac{\frac{g_{n,d\left(n\right)}P_{n}}{N_{0}}}{\frac{g_{n,d\left(n\right)}P_{n}}{N_{0}}+1}$.
For every $k\in\mathcal{K}_{g,n}\left(\mathbf{a}_{-n}\right)$ we
have
\begin{equation}
\underset{n\in\mathcal{N}}{\max}\left|u_{n}\left(k,\mathbf{a}_{-n}\right)-\log_{2}\left(1+\frac{g_{n,d(n)}P_{n}}{N_{0}}\right)\right|\underset{(a)}{\leq}\underset{n\in\mathcal{N}}{\max}\frac{\gamma I_{n}\left(k,\mathbf{a}_{-n}\right)}{N_{0}\ln\left(2\right)}\underset{(b)}{\leq}\frac{\gamma\lambda^{\frac{\alpha}{2}}Gql_{N}}{\ln\left(2\right)\left(\frac{1}{2\theta_{R}}\frac{q-1}{ql_{N}-1}\right)^{\alpha/2}}\frac{P_{0}}{N_{0}}\left(\frac{\log N}{N}\right)^{\frac{\alpha}{2}}\label{eq:10}
\end{equation}
where (a) follows from Lemma \ref{loglog} and (b) from \eqref{eq:9-1}.
Since $BR_{n}\left(\mathbf{a}\right)$ (the exact BR) has better utility
than any of $k\in\mathcal{K}_{g,n}\left(\mathbf{a}_{-n}\right)$,
\eqref{eq:13-3} follows immediately. Since$\frac{l_{N}}{\left(\frac{q-1}{ql_{N}-1}\right)^{\alpha/2}}\left(\frac{\log N}{N}\right)^{\frac{\alpha}{2}}\rightarrow0$
as $N\rightarrow\infty$, we conclude that for a large enough $N$
(such that \eqref{eq:10} is smaller than $\varepsilon+\log_{2}\left(1+\frac{g_{n,d(n)}P_{n}}{N_{0}}\right)-u\left(BR_{n}\left(\mathbf{a}\right)\right)$)
we obtain that each $k\in\mathcal{K}_{g,n}\left(\mathbf{a}_{-n}\right)$
is in $B_{\varepsilon}\left(\mathbf{a}_{-n}\right)$, i.e., $\mathcal{K}_{g,n}\left(\mathbf{a}_{-n}\right)\subseteq B_{\varepsilon}\left(\mathbf{a}_{-n}\right)$
for each $n$ and $\mathbf{a}$. This fact together with \eqref{eq:7},\eqref{eq:8-1}
yield $\underset{N\rightarrow\infty}{\lim}\Pr\left(\forall\mathbf{a},\,\underset{n\in\mathcal{N}}{\min}\left|B_{\varepsilon}\left(\mathbf{a}_{-n}\right)\right|\geq\frac{1}{2l_{N}}\left(1-\frac{1}{q}\right)N\right)=1$.
Note that our analysis is the same regardless of the strategy profile
(hence also $t$) and depends solely on the players' locations.
\end{IEEEproof}

\section{Proof of Lemma \ref{Steps are good in average}}
\begin{IEEEproof}
Denote $N_{\textrm{near},n}=\left|\mathcal{N}_{\textrm{near},n}\right|$
and $\rho=\left(\frac{\log N}{N}\right)^{\frac{\alpha}{2\alpha+4}}$
(see Definition \ref{NearFarSets}). Let $\varepsilon>0$. By the
definition of the approximate BR dynamics, for each $n$ and $\mathbf{a}_{t}$
such that $\mathbf{a_{\mathbf{t}+1}}=BR_{\varepsilon,n}\left(\mathbf{a}_{\mathbf{t}}\right)\neq\mathbf{a}_{t}$
\begin{equation}
E_{n}\left\{ R_{n}\left(\mathbf{a_{\mathbf{t}+1}}\right)-R_{n}\left(\mathbf{a_{\mathbf{t}}}\right)\,|\,\mathbf{a_{\mathbf{t}}}\right\} \geq\frac{\varepsilon}{2}\label{eq:14}
\end{equation}
with probability 1 with respect to the random game. Denote the rate
decrease caused to player $m$ by the fact that player $n$ moves
to his channel as
\begin{equation}
\triangle R_{m}\left(\mathbf{a_{\mathbf{t}+1}}\right)=\triangle R_{m}\left(a_{m},\mathbf{a}_{-n,t}\right)=R\left(I_{m}\left(\mathbf{a_{\mathbf{t}}}\right)+g_{n,d\left(m\right)}P_{n}\right)-R\left(I_{m}\left(\mathbf{a_{\mathbf{t}}}\right)\right)\label{eq:15}
\end{equation}
where $I_{m}\left(\mathbf{a_{\mathbf{t}}}\right)$ is his current
interference, and $g_{n,d\left(m\right)}P_{n}$ is the additional
interference caused by player $n$. For each $m$ such that $m\notin\mathcal{N}_{\textrm{near},n}$
and $d\left(m\right)\neq n$ we have
\begin{equation}
g_{n,d\left(m\right)}\underset{(a)}{\leq}\frac{G}{r_{n,d\left(m\right)}^{\alpha}}\underset{(b)}{\leq}\frac{\pi^{\frac{\alpha}{2}}\lambda^{\frac{\alpha}{2}}G}{\rho^{\alpha}}\label{eq:13}
\end{equation}
where (a) and (b) are for the worst case where the indicator functions
in \eqref{eq:1-1} are equal to 1 and $r_{n,d\left(m\right)}\geq\frac{\rho}{\sqrt{\pi\lambda}}$,
otherwise $g_{n,d\left(m\right)}=0$. Denote $\gamma=\underset{n\in\mathcal{N}}{\max}\frac{\frac{g_{n,d\left(n\right)}P_{n}}{N_{0}}}{\frac{g_{n,d\left(n\right)}P_{n}}{N_{0}}+1}$
and observe that
\begin{equation}
\triangle R_{m}\left(a_{m},\mathbf{a}_{-n,t}\right)\underset{(a)}{\geq}-\frac{\gamma}{\ln\left(2\right)}\frac{I_{m}\left(a_{m},\mathbf{a}_{-n,t}\right)-I_{m}\left(\mathbf{a}_{t}\right)}{N_{0}+I_{m}\left(\mathbf{a}_{t}\right)}\underset{(b)}{\geq}-\frac{\gamma}{\ln\left(2\right)}\frac{g_{n,d\left(m\right)}P_{n}}{N_{0}}\underset{(c)}{\geq}-\frac{\gamma\pi^{\frac{\alpha}{2}}\lambda^{\frac{\alpha}{2}}G}{\ln\left(2\right)\rho^{\alpha}}\frac{P_{0}}{N_{0}}\left(\frac{\log N}{N}\right)^{\frac{\alpha}{2}}\label{eq:22-2}
\end{equation}
where (a) is from \eqref{eq:2}, \eqref{eq:15} and Lemma \ref{loglog},
(b) follows from $I_{m}\left(\mathbf{a}_{t}\right)\geq0$ and (c)
is from \eqref{eq:13} together with $P_{n}\leq P_{0}\left(\frac{\log N}{N}\right)^{\frac{\alpha}{2}}$.

For each $m\in\mathcal{N}_{\textrm{near},n}$ and also for each $m$
such that $d\left(m\right)=n$ we have
\begin{equation}
\triangle R_{m}\left(a_{m},\mathbf{a}_{-n,t}\right)\geq-\underset{m\in\mathcal{N}}{\max\,}R\left(I_{m}\left(\mathbf{a_{\mathbf{t}}}\right)\right)\geq-\underset{m\in\mathcal{N}}{\max}\log_{2}\left(1+\frac{g_{m,d(m)}P_{m}}{N_{0}}\right)\label{eq:22-3}
\end{equation}
Denote $C_{\textrm{max}}=\underset{m\in\mathcal{N}}{\max}\log_{2}\left(1+\frac{g_{m,d\left(m\right)}P_{m}}{N_{0}}\right)$
and $\mathcal{N}_{\textrm{bad},n}=\mathcal{N}_{\textrm{near},n}\bigcup\left\{ m|\,m\in\mathcal{N}\,,d\left(m\right)=n\right\} $\footnote{Player $n$ as a transmitter is ``near'' himself as a destination
for other players (as many as $S$). We take into account the possibility
that player $n$ cannot cancel his own transmission, so $g_{n,n}$
is infinite.}. If indeed $\left|B_{\frac{\varepsilon}{2}}\left(\mathbf{a}_{-n,t}\right)\right|\geq\frac{1}{2l_{N}}\left(1-\frac{1}{q}\right)N$
for some $q>1$ and $N_{\textrm{near},n}\leq\frac{\theta_{T}}{\pi}\rho^{2}N$
then for each $n$ and $\mathbf{a}_{t}$ such that $\mathbf{a_{\mathbf{t}+1}}=BR_{\varepsilon,n}\left(\mathbf{a}_{\mathbf{t}}\right)\neq\mathbf{a}_{t}$
we obtain
\begin{multline}
E_{n}\Biggl\{\sum\limits _{\left\{ m|a_{n,t+1}=a_{m,t}\right\} }\triangle R_{m}\left(a_{n,t+1},\mathbf{a}_{-n,t}\right)\,|\,\mathbf{a_{\mathbf{t}}}\Biggr\}\underset{(a)}{=}\frac{1}{\left|B_{\frac{\varepsilon}{2}}\right|}\sum\limits _{k\in B_{\frac{\varepsilon}{2}}}\sum\limits _{m|a_{m,t}=k}\triangle R_{m}\left(k,\mathbf{a}_{-n,t}\right)\underset{(b)}{=}\\
\frac{1}{\left|B_{\frac{\varepsilon}{2}}\right|}\sum\limits _{m|a_{m,t}\in B_{\frac{\varepsilon}{2}}}\triangle R_{m}\left(a_{m},\mathbf{a}_{-n,t}\right)\geq\frac{1}{\left|B_{\frac{\varepsilon}{2}}\right|}\sum\limits _{m\in\mathcal{N}}\triangle R_{m}\left(a_{m},\mathbf{a}_{-n,t}\right)=\\
\frac{1}{\left|B_{\frac{\varepsilon}{2}}\right|}\sum\limits _{m\notin\mathcal{N}_{\textrm{bad},n}}\triangle R_{m}\left(a_{m},\mathbf{a}_{-n,t}\right)+\frac{1}{\left|B_{\frac{\varepsilon}{2}}\right|}\sum\limits _{m\in\mathcal{N}_{\textrm{bad},n}}\triangle R_{m}\left(a_{m},\mathbf{a}_{-n,t}\right)\underset{(c)}{\geq}\\
-\frac{1}{\left|B_{\frac{\varepsilon}{2}}\right|}\left(\frac{\gamma\pi^{\frac{\alpha}{2}}\lambda^{\frac{\alpha}{2}}G}{\ln\left(2\right)\rho^{\alpha}}\frac{P_{n}}{N_{0}}\sum\limits _{m\notin\mathcal{N}_{\textrm{bad},n}}1+C_{\textrm{max}}\sum\limits _{m\in\mathcal{N}_{\textrm{bad},n}}1\right)\geq-\frac{1}{\left|B_{\frac{\varepsilon}{2}}\right|}\left(\frac{\gamma\pi^{\frac{\alpha}{2}}\lambda^{\frac{\alpha}{2}}G}{\ln\left(2\right)\rho^{\alpha}}\frac{P_{n}}{N_{0}}N+C_{\textrm{max}}\left(N_{\textrm{near},n}+S\right)\right)\underset{(d)}{\geq}\\
-\frac{2l_{N}}{1-\frac{1}{q}}\Biggl(\frac{P_{0}}{N_{0}}\frac{\gamma\pi^{\frac{\alpha}{2}}\lambda^{\frac{\alpha}{2}}G}{\ln\left(2\right)}\rho^{-\alpha}\left(\frac{\log N}{N}\right)^{\frac{\alpha}{2}}+C_{\textrm{max}}\left(\frac{\theta_{T}}{\pi}\rho^{2}+\frac{S}{N}\right)\Biggr)\label{eq:22-1}
\end{multline}
where we omitted $\mathbf{a}_{-n,t}$ from $B_{\frac{\varepsilon}{2}}\left(\mathbf{a}_{-n,t}\right)$
for convenience. Equality (a) follows from the expectation definition
and (b) from changing the order of summation. Inequality (c) follows
from \eqref{eq:22-2} and \eqref{eq:22-3} and (d) from $\left|B_{\frac{\varepsilon}{2}}\left(\mathbf{a}_{-n,t}\right)\right|\geq\frac{1}{2l_{N}}\left(1-\frac{1}{q}\right)N$
and $N_{\textrm{near},n}\leq\frac{\theta_{T}}{\pi}\rho^{2}N$.

By Lemma \ref{NearFar} and Lemma \ref{There are many good actions},
$\underset{n\in\mathcal{N}}{\max}\,N_{\textrm{near},n}\leq\frac{\theta_{T}}{\pi}\rho^{2}N$
and $\underset{n\in\mathcal{N}}{\min}\left|B_{\frac{\varepsilon}{2}}\left(\mathbf{a}_{-n,t}\right)\right|\geq\frac{1}{2l_{N}}\left(1-\frac{1}{q}\right)N$
occur with probability that approaches 1 as $N\rightarrow\infty$.
This implies that inequality (d) in \eqref{eq:22-1} and hence all
\eqref{eq:22-1} is true for each $n$ with probability that approaches
1 as $N\rightarrow\infty$. From \eqref{eq:22-1} it is clear that
since $\frac{P_{0}}{N_{0}}\frac{\gamma\pi^{\frac{\alpha}{2}}\lambda^{\frac{\alpha}{2}}G}{\ln\left(2\right)}\left(\frac{\log N}{N}\right)^{\frac{\alpha}{2}}$
is vanishing, the optimal value of $\rho$ that minimizes the expression
within the brackets is also vanishing to lower the term $\frac{\theta_{T}}{\pi}\rho^{2}C_{\textrm{max}}$.
This accounts for the choice of $\rho=\left(\frac{\log N}{N}\right)^{\frac{\alpha}{2\alpha+4}}$.
Substituting it and adding \eqref{eq:14} and \eqref{eq:22-1} we
get that the probability that for every $\mathbf{a}_{t}$ and $n$
such that $\mathbf{a_{\mathbf{t}+1}}=BR_{\varepsilon,n}\left(\mathbf{a}_{\mathbf{t}}\right)\neq\mathbf{a}_{t}$,
and all $t\geq0$
\begin{equation}
E_{n}\left\{ \varDelta_{1}\left(\mathbf{a_{t}}\right)+\varDelta_{2}\left(\mathbf{a_{t}}\right)|\,\mathbf{a_{\mathbf{t}}}\right\} \geq\frac{\varepsilon}{2}-\frac{2l_{N}}{1-\frac{1}{q}}\left(\frac{\log N}{N}\right)^{\frac{\alpha}{\alpha+2}}\Biggl(\frac{P_{0}}{N_{0}}\frac{\gamma\pi^{\frac{\alpha}{2}}\lambda^{\frac{\alpha}{2}}G}{\ln\left(2\right)}+C_{\textrm{max}}\left(\frac{\theta_{T}}{\pi}+\frac{S}{N^{\frac{2}{\alpha+2}}}\right)\Biggr)>0\label{eq:15-1}
\end{equation}
approaches 1 as $N\rightarrow\infty$.
\end{IEEEproof}
\begin{IEEEbiographynophoto}{}
\bibliographystyle{IEEEtran}
\bibliography{convergence}
\end{IEEEbiographynophoto}

\end{document}